\def\xb{4U1915-05}
\def\ctsspcu{cts s$^{-1}$ PCU$^{-1}$}
\def\ergs{ergs s$^{-1}$}
\def\sa{S$_a$}
\def\ch{$\chi^2$}
\def\chdof{$\rm \chi^2/d.o.f$}
\def\pdetect{P$_{detect}$}
\def\ntrial{N$_{trial}$}

\documentclass{aa}
\usepackage{aabib99,aas_macros,dcolumn,multirow,graphics}
\begin{document}

\thesaurus{06 
(08.14.1;    08.02.3   10.07.2;  13.25.3;    13.25.5;    13.07.2)}   
\title{ Low and High Frequency Quasi-Periodic Oscillations in 4U1915-05}

\author{ L. Boirin\inst{1}, D. Barret\inst{1}, J.F. Olive\inst{1},
  P.F.  Bloser\inst{2}, J.E.  Grindlay\inst{2}} 

\offprints{L. Boirin}

\institute{Centre d'Etude Spatiale des Rayonnements, CNRS/UPS,  9
Avenue du  Colonel Roche,  31028  Toulouse  Cedex 04,  France  (email:
boirin@cesr.fr) \and Harvard  Smithsonian Center for  Astrophysics, 60
Garden  Street,  Cambridge, MA   02138,  USA } 

\date{Received ; accepted}

\titlerunning{RXTE observation of 4U1915-05}
\authorrunning{Boirin et al.}

\maketitle

\begin{abstract} 
  
  The X-ray burster and dipper 4U1915-05 (also known as XB1916-053)
  was observed 19 times for a total exposure of roughly 140 ks between
  1996 February and October with the proportional counter array (PCA)
  aboard {\it Rossi X-ray Timing Explorer} (RXTE).  We present here
  the detailed timing study of its X-ray emission. The source was
  observed in both a high intensity/soft spectral and a low
  intensity/hard spectral state.  The 2-20~keV luminosities are about
  $1.4\times10^{37}$ and $3.2\times10^{36}$~\ergs~in the two regimes
  respectively (assuming a source distance of 9.3 kpc).  We confirm
  that 4U1915-05 is an atoll source based on its spectral behaviour
  and its aperiodic variability in the frequency range
  $\sim4\times10^{-3}-128$~Hz. \xb~displays red noise below
  $\sim$~1~Hz, but is especially remarkable for its quasi-periodic
  variability detected for the first time at several timescales. We
  detect low frequency quasi-periodic oscillations (LFQPOs) between
  $\sim$~5 and 80~Hz, as well as high frequency QPOs (HFQPOs) between
  $\sim$~200 and 1300~Hz.  Both LFQPOs and HFQPOs frequencies
  positively correlate with the mass accretion rate inferred from the
  position of the source on the color-color diagram.  In a narrow
  range of luminosity, we also detect twin simultaneous HFQPOs
  separated by $348\pm12$~Hz.

\end{abstract}

%

\section{Introduction}

\subsection{High frequency quasi-periodic oscillations in low-mass X-ray binaries}

One of the greatest successes of RXTE is the discovery of HFQPOs in
the X-ray emission of low-mass X-ray binaries (LMXBs).  RXTE made it
possible by the combination of the large effective area of the PCA, an
excellent time resolution, an extended telemetry bandwidth and
optimized observing efficiency.  HFQPOs ranging from $\sim300$ to
$\sim1200$~Hz have been reported so far from 21 neutron star LMXBs
\cite{vdk99}: the 6 Z sources, 12 known atoll sources, the unknown
bursting source in the galactic center region, the recently discovered
X-ray bursters XTE J2123-058 and XTE J1723-376
\cite{1723:marshall99iaua}. Following their discoveries, it has been
realized that they could possibly be used to constrain the fundamental
parameters of neutron stars, such as their spin periods, their masses,
and their radii.  Implications of general relativity in the
interpretation of these HFQPOs have also been discussed by several
authors like, e.g., Kaaret et al. \cite*{kaaret97apjl}, Klu\'zniak
\cite*{kluzniak98apjl}, Miller et al. \cite*{miller98conf}.

In most systems, twin HFQPOs are observed with a separation in the
range 220-360~Hz.  Changes in the source luminosity cause the twin
peaks to shift simultaneously in a way that their separation remains
approximatively constant. For the Z source Sco X-1 and the atoll
sources 4U1608-52, 4U1728-34 and 4U1735-44, however, the frequency
separation significantly decreases as the HFQPOs frequencies increase
\cite{scox1:vdk97apjl,1608:mendez98apjlb,1728:mendez99apjl,1735:ford98apjl,psaltis98apjl}.
In a few X-ray bursters (e.g. 4U1702-43, KS1731-260), which are atoll
sources, coherent pulsations have been seen during type I X-ray bursts
at frequencies equal or twice the frequency difference between the
twin HFQPOs \cite{1702:markwardt99apjl,1731:smith97apjl}.  In
4U1636-53 and 4U1728-34, burst oscillations have been detected at a
frequency significantly greater than the twin HFQPOs separation
\cite{1636:mendez98apjl,1728:mendez99apjl}.

\begin{table*}[!ht]
\begin{center}
\scriptsize
\renewcommand{\arraystretch}{1}
\begin{tabular}{llclcl}
\hline
Source &  Type$^{\rm a}$ & \multicolumn{2}{c}{Flat topped noise and
  QPO} & \multicolumn{2}{c}{Red noise and QPO} \\
 & &  & References$^{\rm b}$ & & References$^{\rm b}$\\
\hline
\hline
4U 0614+09 & A & Y & 1$^{\rm c}$, 2 & Y & 44$^{\rm c}$, 45 \\
4U 1608-52 & A & Y & 3, 4, 5, 2 & - &\\
4U 1702-42 & A & - & & Y & 14$^{\rm c}$ \\
4U 1705-44 & A & Y & 6, 5, 7 & Y & 5$^{\rm c}$\\
1E 1724-30 & A & Y & 8, 9 & - & \\
4U 1728-34 & A & Y & 10, 11 &  - &\\
KS 1731-260 & A & - & & Y & 15, 9 \\
4U 1735-44 & A & Y & 2 & Y & 16\\
SLX 1735-269 & A & Y & 12, 9 & - & \\
4U 1744-26 & A & - & & Y & 17, 18 \\
XTE J1806-246  & -$^{\rm d}$ & - & & Y & 19, 20 \\
SAX J1808.4-3658  & P & Y & 13 & - & \\
GX 13+1 & A & - & & Y & 21 \\
4U 1812-12 & A & Y & 2 &- &  \\
4U 1820-30 & A & - & &Y &  22 \\
GS 1826-238 & A & Y & 9 &- & \\
4U 1915-05 & A & - & & Y & 46\\
XTE J2123-058 & A &- & & Y & 23, 24\\
Sco X-1 & Z &Y &  25, 2 & Y & 25, 26, 27\\
GX 340+0 & Z & Y & 2 &  Y & 28\\
GX 349+2  & Z & - & & Y & 29\\
GX 5-1 & Z & Y & 2 & Y & 30, 31, 32\\
GX 17+2 & Z &  Y & 2, 33 &  Y & 33, 34, 35\\
Cyg X-2 & Z &  Y & 2, 36, 37, 38 &  Y &36, 37, 38, 39\\
Cir X-1 & Z$^{\rm e}$ &  Y & 40, 41, 42, 43 & Y & 40, 42, 43 \\
\hline
\end{tabular}\\
\end{center}
$^{\rm a}$ A, atoll source; Z, Z source; P, millisecond X-ray
pulsar. \\
$^{\rm b}$ See  Van Paradijs \cite*{vp95xrb} for complementary
references up to 1995.\\
$^{\rm c}$ After the power spectra published in the literature.\\
$^{\rm d}$ Both atoll and Z classifications have been proposed
\cite{1806:wijnands99apj,1806:revnivtsev99aa}.\\
$^{\rm e}$ Shirey et al. \cite*{cirx1:shirey98apj,cirx1:shirey99apja}\\
 References: 
 (1) Méndez et al. \cite*{0614:mendez97apjl};
 (2) Wijnands and Van der klis \cite*{wijnands99apj};
 (3) Yoshida et al. \cite*{1608:yoshida93pasj};
 (4) Yu et al. \cite*{1608:yu97apjl};
 (5) Berger and Van der Klis \cite*{1705:berger98aa};
 (6) Hasinger and Van der Klis \cite*{hasinger89aa};
 (7) Ford et al. \cite*{1705:ford98apjl};
 (8) Olive et al. \cite*{tz2:olive98aa};
 (9) Barret et al. \cite*{barret00apj};
 (10) Strohmayer et al. \cite*{1728:stro96apjl};
 (11) Ford and Van der Klis \cite*{1728:ford98apjl};
 (12) Wijnands and Van der Klis \cite*{slx:wijnands99apjl};
 (13) Wijnands and Van der Klis \cite*{1808:wijnands98apjl};
 (14) Markwardt et al. \cite*{1702:markwardt99apjl};
 (15) Wijnands and Van der Klis \cite*{1731:wijnands97apjl};
 (16) Wijnands et al. \cite*{1735:wijnands98apjl};
 (17) Lewin et al. \cite*{1744:lewin87mnras};
 (18) Strohmayer \cite*{1744:stro98conf};
 (19) Wijnands and Van der Klis \cite*{1806:wijnands99apj};
 (20) Revnivtsev et al. \cite*{1806:revnivtsev99aa};
 (21) Homan et al. \cite*{1811:homan98apjl};
 (22) Wijnands et al. \cite*{1820:wijnands99apjl};
 (23) Homan et al. \cite*{2123:homan99apjl};
 (24) Tomsick et al. \cite*{2123:tomsick99apj};
 (25) Van der Klis et al. \cite*{scox1:vdk97apjl};
 (26) Van der Klis et al. \cite*{scox1:vdk96apjl};
 (27) Dieters and Van der Klis \cite*{scox1:dieters00mnras};
 (28) Jonker et al. \cite*{gx340:jonker98apjl};
 (29) Kuulkers and Van der Klis \cite*{gx349:kuulkers98aa};
 (30) Kamado et al. \cite*{gx5:kamado97pasj};
 (31) Wijnands et al. \cite*{gx5:wijnands98apjl};
 (32) Vaughan et al. \cite*{gx5:vaughan99aa};
 (33) Kuulkers et al. \cite*{gx17:kuulkers97mnras};
 (34) Wijnands et al. \cite*{gx17:wijnands96apjl};
 (35) Wijnands et al. \cite*{gx17:wijnands97apjl};
 (36) Wijnands et al. \cite*{cygx2:wijnands97aa};
 (37) Kuulkers et al. \cite*{cygx2:kuulkers99mnras};
 (38) Focke \cite*{cygx2:focke96apjl};
 (39) Wijnands et al. \cite*{cygx2:wijnands98apjl};
 (40) Oosterbroek et al. \cite*{cirx1:oosterbroek95aa};
 (41) Shirey et al. \cite*{cirx1:shirey96apjl};
 (42) Shirey et al. \cite*{cirx1:shirey98apj};
 (43) Shirey et al. \cite*{cirx1:shirey99apja};
 (44) Ford et al. \cite*{0614:ford97apjla};
 (45) Ford \cite*{ford97thesis};
 (46) This paper.\\

\caption{Sources for which a QPO (or a QPO-like feature or peaked HFN) 
  has been reported in addition to either
  flat topped noise or red noise in their low frequency
  ($\sim10^{-3}-100$~Hz) PDS.}
\label{qporeview}
\end{table*}

Specific also to the atoll sources, HFQPOs are detected when the
source is in the so-called island state and on the lower branch of the
banana state. These two states are easily identified in color-color
diagrams and there is conclusive evidence that the accretion rate
increases from the island to the banana state \cite{hasinger89aa}.
They are associated with different types of power density spectra
(PDS) at low frequencies (in the remainder of this paper we define low
frequencies as frequencies below 100~Hz). In the island state, the
source displays high frequency noise (HFN also called ``flat topped
noise'') and the PDS can be approximated by a zero-centered Lorentzian
\cite[for 1E1724-3045]{tz2:olive98aa}, or alternatively by broken
power laws \cite[for example]{wijnands99apj}.  Superposed to the
noise, a QPO-like feature is generally seen around $\sim0.5-70$~Hz
(see Table \ref{qporeview} for a review).  In this state, the root
mean square (RMS) values of the fluctuations (to the mean flux) reach
50~\% \cite{vdk94apjs}. On the other hand, on the banana branch, the
shape of the PDS is roughly a power law with an index of -1.0; the
so-called very low frequency noise (VLFN) or red noise
\cite{hasinger89aa}.  In this state, the RMS is around 10~\% or less.
In the lower part of the banana branch, a broad structure, often
refered to as ``peaked HFN'', or a QPO is seen at frequencies in the
range 1-70~Hz (see Table \ref{qporeview}). A correlation between the
centroid frequency of the bump, and the frequency of the upper kHz QPO
has been found in several LMXBs (e.g.  KS1731-260), and interpreted in
the framework of a Lense-Thirring effect
\cite{stella98apjl,stella99apjl} or a two-oscillator model
\cite{tita99apjl,1702:osherovich99apjl}.

\subsection{\xb}

\begin{table*}[!t]
\begin{center}
\scriptsize
\renewcommand{\arraystretch}{1}
\begin{tabular}{llllllllllll}
\hline
Name & Start & & End & &  P & M &  T$_{\rm total}$  & B & D & T  & R \\
\hline
\hline 
02/10$^{(1)}$  & Feb. 10 &  00:14:04 & Feb. 10 & 05:09:13 & 5 & G & 6750 & 0 & 3 & 4112  & 73.0 \\ 
03/13$^{(1)}$ & March 13  & 22:41:24 & March 14  &00:54:13 & 5 & G & 3316  & 0 & 0 &2880 & 97.1 \\
05/05 & May 5  & 05:59:39 & May 5 & 09:46:13 & 3, 4 & G & 4039 & 0 & 0 & 3904 & 68.7 \\ 
05/06 & May 5  & 22:16:00 & May 6  &03:37:13 & 4 & G, CB & 6912 & 1 & 0 & 6592 &  62.1 \\ 
05/14 & May 14 & 09:22:05 & May 14 & 12:55:13 & 3 & E & 7599 & 0 & 3 & 4432 & 29.5 \\  
05/15 & May 15 & 12:32:06 & May 15 & 15:11:13 & 3 & E & 6622 & 0 & 0 & 6592 & 41.5 \\
05/16 & May 16 & 12:56:45 & May 16 & 16:31:13 & 3 & E & 7955 & 0 & 4 & 5936 & 42.6 \\ 
05/17 & May 17 & 08:20:08 & May 17 & 11:49:13 & 3 & G & 7640 & 0 & 4 &2864 & 35.4 \\ 
05/18 & May 18 & 09:53:47 & May 18 & 13:20:13 & 3 & E & 7495 & 0 & 4 & 3168 &21.7 \\
05/19 & May 19 & 10:06:38 & May 19 & 13:37:13 & 3 & E & 7389 & 0 & 2 & 2560 & 22.6 \\ 
05/20 & May 20 & 06:50:10 & May 20 & 10:16:13 & 3 & E & 7496 & 0 & 2 & 3904 & 25.2 \\ 
05/21 & May 21 & 07:47:52 & May 21 & 11:22:13 & 3 & E & 7949 & 0 & 4 & 2400 & 23.3 \\ 
05/22 & May 22 & 12:37:55 & May 22 & 16:26:13 & 3 & E & 8624 & 0 & 6 & 3584 & 27.6 \\ 
05/23$^{a}$ & May 23 & 07:50:56 & May 23 & 11:20:13 & 4, 5 & E & 7723 & 0 & 8 &
4128 & 39.8 \\
06/01 & June 1  &17:36:53 & June 1 & 21:36:13 & 5 &E, G, CB& 9336 & 1 & 4 & 4544 & 46.3 \\ 
07/15 & July 15 & 11:49:45 & July 15 & 16:24:13 & 5 & E & 9786 & 0 & 6 & 5168 & 33.9 \\ 
08/16 & Aug. 16  &10:37:49 & Aug. 16 &  14:55:13 & 4, 5 & E & 9983 & 1 &5& 6384& 28.1 \\
09/06 & Sept.  6 & 15:57:13 & Sept. 6 & 21:26:13 & 5 & E & 9344 & 0 & 1 &6032 & 39.5 \\ 
10/29 & Oct. 29 & 06:53:21 & Oct. 29 & 11:53:13 & 5 & E & 9464 & 1 & 2 & 5120 & 35.9 \\ 
\hline
\end{tabular}
\end{center}
\caption{Observation log. 
We list the name assigned to the observation, its start and stop
times (day and UT hour), the number of PCUs working (P), the high time
resolution data modes M available (G, E and CB designate  the
E\_125us\_64M\_0\_1s, GoodXenon\_16s and CB\_2ms\_64M\_0\_249 (Binned
Burst Catcher) configurations respectively),
the total  exposure   time (T$_{\rm total}$) in seconds after screening (on
elevation and offset), the number of bursts (B) and  dips (D) present, the
exposure time in seconds 
of the persistent light curve after filtering out the
bursting and dipping parts (T)  and the 2-20~keV background subtracted count
rate in units of~\ctsspcu~(R) of the persistent emission. $^{(1)}$ Gain epoch 1
data (the other observations are gain epoch 3). $^{a}$ Two small
observations were carried out on May 23rd, we merged them together for 
clarity. } 
\label{exposure}
\end{table*}

In the context of these results, we have analyzed our RXTE
observations of the type I X-ray burster \xb. This source is known as
the first dipping X-ray source discovered, providing the first
reliable evidence for the binary nature of such sources
\cite{1916:walter82apjl,1916:white82apjl,1916:swank84apj,1916:smale88mnras}.

The period of the dips in X-rays is 50 minutes, the shortest among
dipping sources
\cite{1916:white82apjl,1916:walter82apjl,1916:yoshida95pasj,1916:chou99aas}.
It is also optically identified with a 21st V magnitude blue star
\cite{1916:grindlay88apjl}. A modulation is seen in the optical but at
a period $\sim$~1~\% larger than the X-ray dip period
\cite{1916:callanan95pasj}.  The discrepancy between the two periods
is confirmed by Chou et al. \cite*{1916:chou99aas} who combined
historical and recent data (including the RXTE data used in this
paper).  This discrepancy has led to several interpretations among
which the SU UMa\footnote{SU UMa stars are cataclysmic variables,
  which during their long (`super') outbursts show humps which have a
  period which is a few \% longer than their orbital period.}  model
is prefered, although the hierarchical triple system model proposed by
Grindlay et al. \cite*{1916:grindlay88apjl} is still possible
\cite{1916:chou99aas}.

To date, the only timing study of the non-dipping and non-bursting
emission of \xb~was carried out with the Ginga data, which represented
the most extensive X-ray observations of the source before the launch
of BeppoSAX and RXTE.  The 1-37~keV luminosity varied between
$5\times10^{36}$ and $1.2\times10^{37}$~\ergs, in the range of
luminosities previously observed for the source.  A V-shape was
observed in the hardness (7-18/1-7~keV) intensity (1-18~keV) diagram
\cite{yoshida92thesis}.  The $3\times10^{-3}-1$~Hz PDS were fitted
with a power law of index -0.6 for the left branch of the V data and
of index roughly -1.1 for the right branch data.  Yoshida
\cite*{yoshida92thesis} suggested that \xb~was an atoll source.
Yoshida \cite*{yoshida92thesis} interpreted the left branch of the
hardness intensity diagram as the island state and the right branch as
the banana state, although the two branches overlapped in the
color-color diagram.

From an X-ray burst showing photospheric radius expansion in the GINGA
data, the distance to the source was estimated as 9.3 kpc
\cite{yoshida92thesis}.  \xb~was observed at energies up to 100~keV by
BeppoSAX \cite{1916:church98aa}. This was the first observation of a
hard X-ray tail in the spectrum of 4U1915-05. Indeed, previous pointed
and monitoring observations at high energies with OSSE and BATSE
aboard the Compton Gamma-Ray Observatory had failed to detect the
burster \cite{1916:barret96aas,1916:bloser96aas}.  \xb~is the first
dipping source to belong to the growing group of LMXBs detected up to
$\sim100$~keV \cite{barret00apj}.

In this paper we report on the timing analysis of the X-ray emission
which led to the discovery of HFQPOs and LFQPOs in \xb. In the next
section, we describe the observations and the data analysis. Then, we
present the results.  Finally, we compare \xb~to similar sources and
discuss the results in the framework of currently proposed models of
HF and LFQPOs.

\section{Observations and data analysis}

\subsection{The RXTE/PCA observations}

\begin{figure}[!t]
\resizebox{\hsize}{!}{\includegraphics{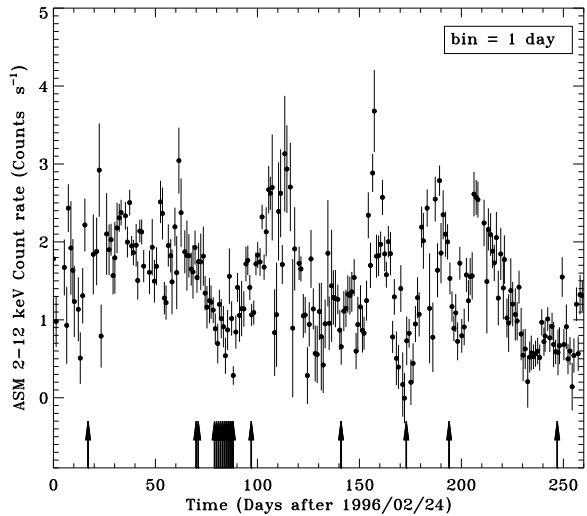}}
\caption{Light curve of \xb~as observed by the RXTE/ASM. The binning
  time is one day. The times of the pointed observations are indicated
  by arrows.}
\label{asmlc}
\end{figure}

\xb~was observed by RXTE on 19 occasions in 1996: on February 10th,
March 13th, May 5-6th, May 14th to 23rd, June 1st, July 15th, August
16th, September 6th and October 29th. The log of the observations is
presented in Table \ref{exposure}.  An observation has an exposure
time of roughly 8000 seconds and is provided as typically three data
files with gaps between files. These files are refered as segments of
observation in our analysis.  We report on the data collected in the
2-60~keV range with the PCA.  It consists of 5 nearly identical large
area Proportional Counter Units (PCU 0 to 4) corresponding to a total
collecting area of $\sim6500$ cm$^2$ \cite{jahoda96}. For safety
reasons, PCUs are switched on and off in the course of an observation.
During our observations, the data were mainly obtained in three
different PCU configurations: all PCUs on, first four on and first
three on.  Furthermore, our observations cover two PCA gain epochs:
the February 10th and March 13th observations belong to the first one
whereas the following observations belong to the third one.  Different
response matrices were used for each set.  More than 138 ks of good
data are available after recommended screening (elevation angle
above the Earth's limb greater than 10 degrees and pointing offset
angle less than 0.02 degree).

The all-sky monitor (ASM) aboard RXTE consists of three coded-aperture
cameras mounted on a motorized rotation drive, allowing to view
different regions of the sky during a satellite orbit
\cite{levine96apjl}. The instrument is sensitive in the energy range
$\sim$~2-12~keV.  Fig. \ref{asmlc} is the ASM light curve of
\xb~during 1996. The binning time is one day. The times of the PCA
observations are flagged with arrows.  Four X-ray bursts were recorded
during the PCA observations: on May 5th, June 1st, August 16th and
October 29th.  With the exception of the first one, they began during
a dip. 
A total of 58 primary or secondary dips (or parts of dips) were
observed, lasting typically 400 seconds. Dips are thought to be due to
occultation of the internal emitting region by vertical structure at
the outer edge of the accretion disk
\cite{1916:walter82apjl,1916:white82apjl}. Primary dips occur at
intervals consistent with an orbital period of $\sim$~50 minutes.
Secondary dips are ``anomalous'' dips, narrower than the primary ones.
They occur irregularly at $\sim$~180$^\circ$ out of phase with the
primary dips \cite{1916:chou99aas}. Dipping is almost 100~\% in the
3-5~keV energy range, but only $\sim$~65~\% in the 5-30~keV range.

\subsection{Data modes}

The {\it Standard 2} mode is available for each observation. This mode
provides counts integrated during 16 seconds in 128 energy channels
covering the 2-60~keV energy range.  The high time resolution data
were provided in two different modes, depending on the observations
(see Table \ref{exposure}): the E\_125s\_64M\_0\_1s mode with
122~$\mu$s resolution or the GoodXenon\_16s mode with $\sim1~\mu$s
resolution.  The bursts that occured on October 29th and August 16th
were recorded in the 122~$\mu$s resolution mode. The bursts that
occured on May 5th and June 1st were recorded during 3.75 seconds in a
burst catcher mode with 2 ms resolution.

\subsection{Data analysis}

\subsubsection{Light curves, color-color and hardness-intensity
  diagrams}

For the light curves, color-color and hardness-intensity diagrams, we
have used the {\it Standard 2} mode data.  We made light curves in
adjacent energy bands.  The bursts and dips were then filtered out
after visual screening of the light curves to generate the color-color
and hardness-intensity diagrams for the persistent emission.  The soft
color was calculated as the 3-5~keV/1.7-3~keV count rate ratio, and
the hard color as 10-30~keV/5-10~keV.

As all the PCUs have different energy responses, the absence of a PCU
can  mimic spectral variations.  Thus, we used the data collected
with the first three PCUs since they were continuously operating
during all our observations.

Differences between response matrix of gain epoch 1 and 3 introduce
differences in the boundaries of the energy channels used. This
instrumental effect introduces shifts between the colors or count
rates obtained from different epochs. We quantified these shifts using
the Crab nebula which is a steady source. We find that the correction
factors to apply to gain epoch 1 data in order to make them match the
epoch 3 region are +10.1~\%, -1.5~\%, -1.1~\% and +2.8~\% for the
count rates in the energy bands 1.7-3, 3-5, 5-10 and 10-30~keV
respectively.  The correction factors are -10.5~\% and +4.0~\% for the
soft and hard colors respectively. We applied these correction factors
to \xb~data.

\subsubsection{Power density spectra}

To investigate the variability of \xb, we computed PDS using the
FTOOLS {\it powspec}. We treated separately the data obtained from the
different high time resolution modes.  
To reach timescales ranging from roughly 10 ms to 4 minutes, the
data were rebinned to a $\sim4$~ms resolution. Then, each continuous
set was divided into segments of 65536 bins. A fast Fourier transform
(FFT) of each segment was computed yielding a low frequency PDS in the
range $\sim4\times10^{-3}-128$~Hz.

To reach timescales ranging from roughly 0.25 ms to 1 second, the data
were rebinned to a $\sim122~\mu$s resolution.  Each continuous set was
divided into segments of 8192 bins. A FFT of each segment was computed
leading to high frequency PDS in the range 1-4096~Hz.  Then, the low
or high frequency PDS obtained from a given set of observations, a
given observation or a given segment of observation were averaged
together to obtain a final PDS representative of the set, observation
or segment.

The PDS were normalized according to Leahy et al. \cite*{leahy83apj}
so that the average power expected from a Poisson distribution is 2.
We have checked in the high frequency PDS that the average white noise
level above 1500~Hz was indeed 2, indicating that deadtime corrections
were not needed.

The error on each PDS bin has been set to $E_P =
\frac{P_{m}}{\sqrt{MW}}$ where M is the number of raw PDS averaged
together, W the number of raw frequency bins averaged together
\cite{vdk89nato}.  $P_{m}$ is set to the Fourier power at the point
considered for low frequency PDS (where the power can differ
significantly from the white noise level 2 because of the source
aperiodic variability) and to 2 for the high frequency PDS.  Both
logarithmic and linear rebinnings were applied.

Different components were summed to fit the resulting PDS: a constant
for the white noise level (C), a power law for the VLFN (PL), a cutoff
power law for the HFN (CPL). The functional shapes used for the latter
components are $\rm A\nu^\alpha$ and $\rm A\nu^\alpha
exp(-\nu/\nu_{\rm cutoff})$ respectively, where A is the
normalization, $\alpha$ the power law index, $\nu$ the frequency and
$\nu_{\rm cutoff}$ the cutoff frequency. C is a free parameter for the
high frequency PDS and is set to 2 for the low frequency PDS. We used
Gaussians (G) to describe the QPO features. Note that Lorentzians are
also currently used \cite{vdk95xrb}.

All the fits were performed using a \ch~minimization technique.  The
errors on the analytic model parameters correspond to a \ch~variation
of 2.7, equivalent to the 90~\% confidence region for a single
interesting parameter.

To compute the RMS, the PDS were normalized according to Belloni and
Hasinger \cite*{belloni90aa} taking into account the total and
background count rates. The RMS was then obtained by integrating the
normalized PDS in the appropriate frequency range.

We have also computed PDS of data from the bursts.  For the bursts
recorded in the 122~$\mu$s resolution mode, we have calculated an FFT
power spectrum using 4096 bins lasting 244~$\mu$s, so that one PDS
corresponds to a segment of the burst of 1 sec duration and the
Nyquist frequency is 2048~Hz.  For the other bursts, the time
resolution is only $\sim$~2 ms, yielding a Nyquist frequency of
256~Hz.

\subsubsection{Search technique for quasi-periodic oscillations}
\label{sec:search_technique}

To look for weak QPOs in the PDS (as in the case of \xb), we developed
an algorithm which computes for  each PDS the width of the window
which maximises the signal to noise ratio (SNR) within the window.
Assuming that a given window of width $w$ contains N frequency bins,
the error on the binned points becomes $E_P = \frac{P_m}{\sqrt{MWN}}$
and the SNR can be computed as:

\begin{equation}
SNR (in\; \sigma) = \frac{\frac{1}{N} \sum P_j - P_{ref}}{E_P}
\end{equation}

where P$_j$ are the N PDS points contained in the window of width $w$
and P$_{ref}$ is the reference power. For the low frequency PDS, given
the weakness of the VLFN component above $\sim$~1~Hz, we looked for
excesses above $P_{ref}$=2, the theoretical level expected from
Poissonian noise.  PDS have been processed through the algorithm and
searched for excesses of widths from 1 to 50~Hz in the frequency range
1-100~Hz.  For the high frequency PDS, we looked for excesses of
widths 5-200~Hz in the frequency range 100-1500~Hz.  We used the noise
level averaged above 1500~Hz ($\sim$~2.0) as the reference $P_{ref}$
for the search of excesses in the PDS.  In both cases, we first
selected the positive detections above 3.5~$\sigma$.  A new
significance level inferred from the fit was then attributed to each
excess in order to take into account the excess shape. We report here
on the LFQPOs and HFQPOs which have a significance level greater than
3~$\sigma$. This significance level is reported in Table \ref{tabqpo}
with the properties of the QPOs detected in \xb. We do mention either
the observation or the segments, so that no overlapping is possible.

This algorithm does not take into account the number of trials in
determining the significance level.  To determine the real
significance of the detections, we follow Van der Klis
\cite*{vdk89nato} and define the $(1-\epsilon)$ confidence level
\pdetect~as the power level that has only the small probability
$\epsilon$/\ntrial~to be exceeded by a noise power.  \ntrial~is the
number of different power values that one wishes to compare with
\pdetect. It may be the total number of bins in the PDS, or less if
only a given frequency range is looked for excesses.  The number of
different PDS looked at may also be included in the number of trials
but we did not take it into consideration. In our case, the high
frequency PDS are scanned over a frequency range of 1400~Hz. The
exposure time of a raw PDS is 1 second (corresponding to a frequency
resolution of 1~Hz). The typical exposure time for a segment of
observation is 3000 seconds so that M=3000 raw PDS are averaged
together to create the PDS representative of the segment.  Assuming
that the algorithm finds a maximum excess of width 20~Hz
(corresponding to a final rebinning factor NW of 20) at 5~$\sigma$,
then \ntrial=70 and the actual significance level of the signal is
99.998~\%.  In the worst case of an excess of width 5~Hz (ie the
minimum width considered) found at 3.5~$\sigma$ (ie the minimum level
obtained from the algorithm considered), \ntrial=280 and the actual
significance level is 92.7~\%.  The low frequency PDS are scanned over
a frequency range of 100~Hz.  The exposure time of a raw PDS is
$\sim$~256 seconds (corresponding to a frequency resolution of
$4\times10^{-3}$~Hz). A PDS representative of a 3000 seconds segment
is obtained from M=12 raw PDS. In the worst case of an excess of
width 1~Hz (e.g NW=256) at 3.5~$\sigma$, then \ntrial=100 and
the true significance level is 97.0~\%.

\section{Results}

\subsection{Light curves, color-color and hardness-intensity diagrams}

\begin{figure}[!b]
\resizebox{\hsize}{!}{\includegraphics{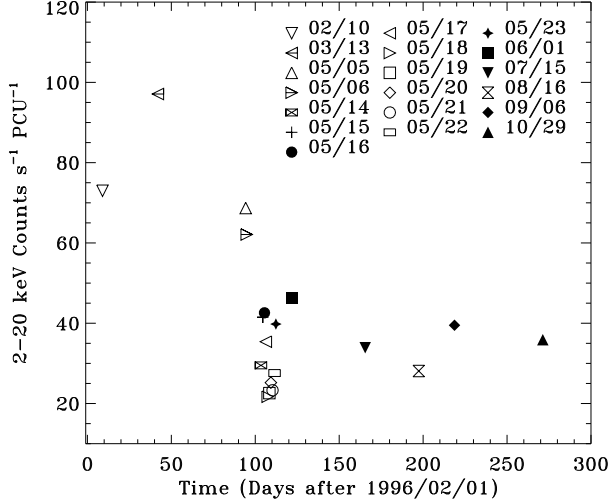}}
\caption{Evolution of the persistent emission during the observations.
  The 2-20~keV background subtracted count rates are averaged over
  each observation. The corresponding exposure time is given in Table
  \ref{exposure}. The symbols labeled with the date of the
  observation will be used throughout the remainder of this paper.}
\label{tot_lc}
\end{figure}
\begin{figure}[!b]
\resizebox{\hsize}{!}{\includegraphics{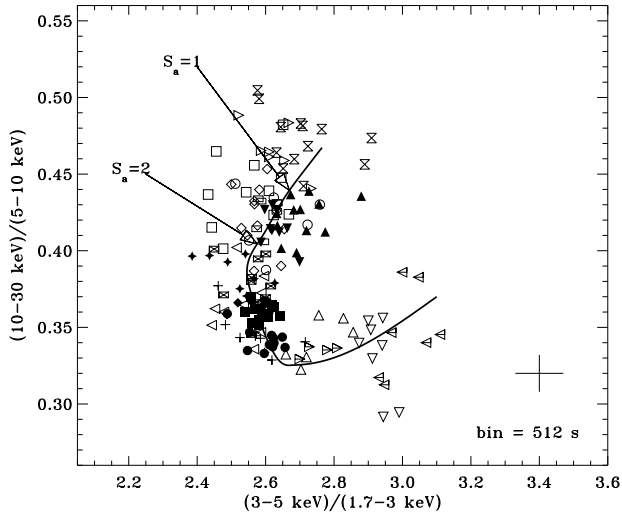}}
\caption{Color-Color diagram. The binning time is
  512 seconds and a typical error bar is shown. The thick line is the
  spline used to approximate the position on the color-color diagram.
  This position is quantified by the curvilinear coordinate S$_a$
  chosen arbitrary. The positions S$_a$=1 and S$_a$=2 are indicated by
  arrows.}
\label{color_color}
\end{figure}
\begin{figure}[!b]
\resizebox{\hsize}{!}{\includegraphics{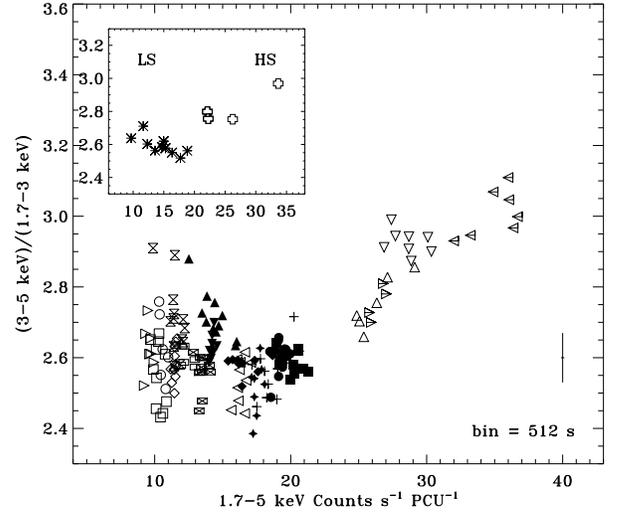}}\\%
\resizebox{\hsize}{!}{\includegraphics{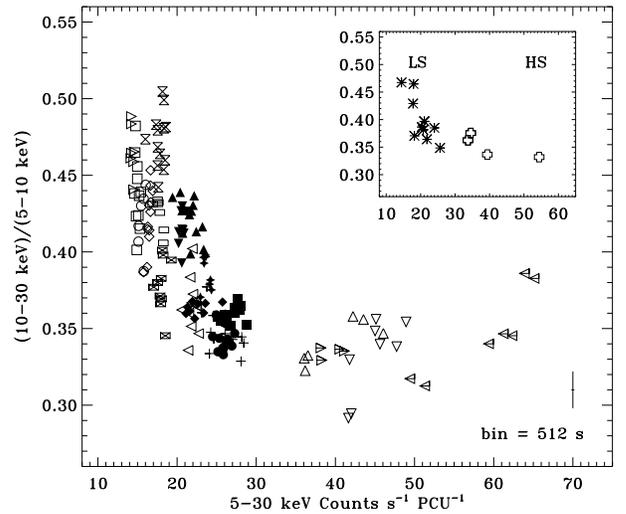}}
\caption{Hardness-intensity diagrams for the soft color (top panel) and 
  the hard color (bottom panel). In the main panels, the binning time
  is 512 seconds and a typical error bar is shown. In the inset
  panels, each point is the average over an observation. The asterisks
  and the large crosses indicate observations of the LS and HS respectively.}
\label{hard_int}
\end{figure}

Fig. \ref{tot_lc} shows the global evolution of the persistent
emission during our observations. Each point gives the 2-20~keV
background subtracted count rate integrated over a single observation.
Each plotting symbol is associated with an observation (see the right
corner of Fig. \ref{tot_lc}).  As could have been anticipated from
Fig. \ref{asmlc}, our observations sampled various intensity states of
the source: the PCA count rate changes by a factor of $\sim$~5 from
roughly 20 to 100~\ctsspcu.  We also generated light curves in
different energy bands between 2 and 30~keV for each observation.  No
substantial variation is noticeable in the persistent emission over
timescales of a few hours.

\begin{figure*}
\resizebox{\hsize}{!}{\includegraphics{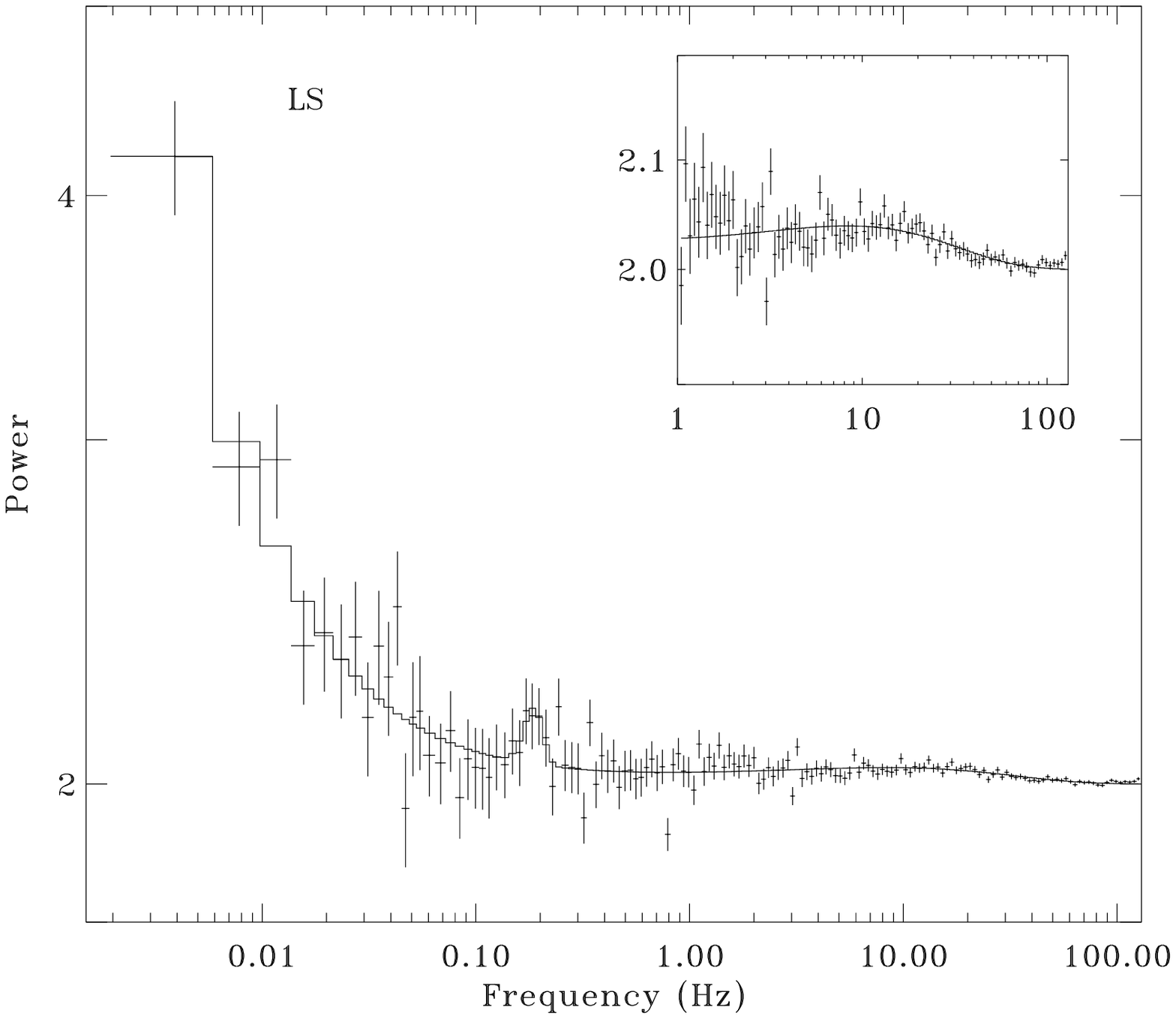}\includegraphics{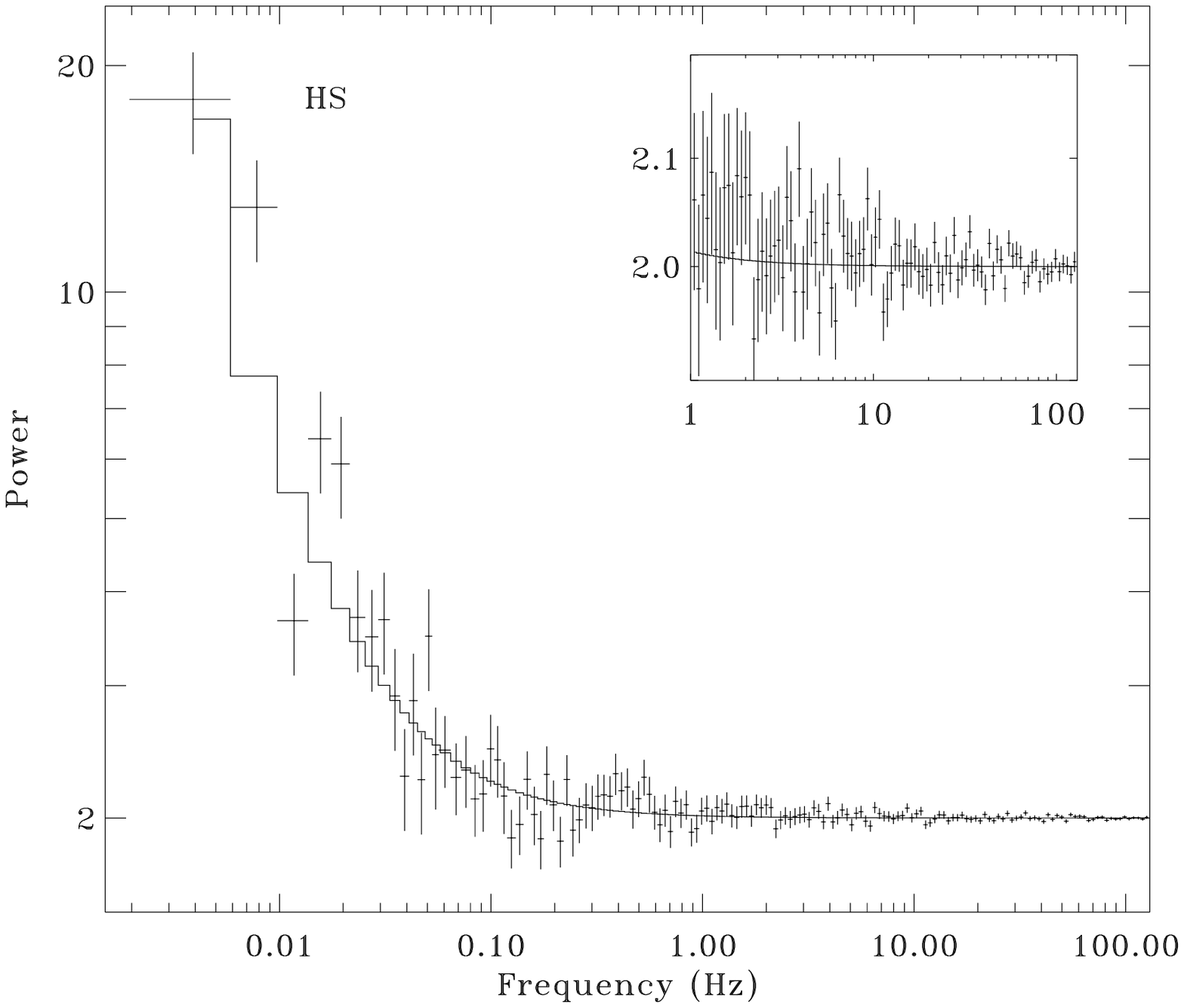}}
\caption{Average PDS in the LS (left panel) and HS (right panel). The
  first one is obtained by averaging the PDS of the LS observations
  available in the 122~$\mu$s resolution mode: all the LS observations
  except a segment of the June 1st observation and the May 17th
  observation. The second PDS is obtained by averaging the PDS of the
  HS observations: February 10th, March 13th, May 5th and 6th (all
  obtained in 1~$\mu$s resolution mode). The total exposure time of
  the first PDS is more than 15~hours whereas it is only
  $\sim$~3~hours for the second one. The inset panel is a zoom of the
  PDS above 1~Hz. The solid lines are the best fits to the data.}
\label{summedpds}
\end{figure*}
\begin{table*}
\begin{center}
\scriptsize
\renewcommand{\arraystretch}{1}
\begin{tabular}{llccccccccc}
\hline
State & Energy & \multicolumn{2}{c}{VLFN} & \multicolumn{3}{c}{HFN} &
\multicolumn{3}{c}{QPO}  &\chdof \\
&  & $\alpha$ & RMS & $\alpha$ & $\nu_{\rm cutoff}$ & RMS
&$\nu$&FWHM&RMS& \\ 
& (keV) & &(\%)&&(Hz)&(\%)& (Hz)&(Hz)&(\%)&\\
\hline
\hline
LS & 5-30 & -1.03$^{+0.16}_{-0.14}$ & 3.7$\pm$0.2 & 0.5$^{+0.2}_{-0.3}$ & 18.9$^{+6.4}_{-4.7}$ & 17.5$\pm$0.5 & 0.19$^{+0.03}_{-0.02}$ & 0.02$^{+0.04}_{-0.01}$ & 1.9$\pm$0.2 & 159/143\\
& 2-30 &  -0.92$\pm$0.08 & 3.8$\pm$0.1 & 0.6$^{+0.2}_{-0.3}$ & 14$\pm$3 & 15.4$\pm$0.3 & 0.19$^{+0.02}_{-0.01}$ & 0.014$^{+0.032}_{-0.004}$ &1.9$\pm$0.2 &175/143\\
HS & 5-30 & -1.24$^{+0.14}_{-0.13}$ & 3.9$\pm$0.2 & -& -& 3.8$\pm$0.6 & -&- & 0.8$\pm$0.3 &  146/149 \\
\hline
\end{tabular}
\end{center}
\caption{Best fit parameters obtained for the state averaged PDS shown in 
  Fig. \ref{summedpds}. The RMS is given in the frequency range 
  $4\times 10^{-3}-1$~Hz for the VLFN, 1-128~Hz for the HFN and 0.1-0.3~Hz for the QPO.}
\label{fit}
\end{table*}

Fig.  \ref{color_color} shows the color-color diagram.  The points
seem to form a ``Banana'' shape than can be approximated by a spline.

Fig.  \ref{hard_int} shows the soft (top panel) and hard (bottom
panel) colors as a function of the count rate.  In the main panels,
the binning time is 512 seconds while in the inset panels, each point
is the average over an observation.  

Looking at the behaviour of the source within an observation, we see,
from the vertical tracks drawn by the points with a same symbol (main
panels), that the colors can vary a lot while the intensity remains
nearly constant, especially the hard color at low count rates.

Looking at the global behaviour of the source along the observations
(inset panels), two groups corresponding to two different spectral
states may be distingished: we refer to the observations with the 5-30
keV intensity below and above $\sim$~30~\ctsspcu~as being in the low
and high states (LS, HS) respectively.  In the LS, we see a trend for
the mean hard color to decrease as the intensity increases (Fig.
\ref{hard_int} bottom), while the soft color remains nearly constant
(Fig. \ref{hard_int} top), so that, on average, the emission softens
with the intensity.  On the contrary, in the HS, the hard color is
nearly constant (Fig.  \ref{hard_int} bottom), while the soft color
increases with the intensity (Fig.  \ref{hard_int} top), so that, on
average, the emission hardens.  \xb~is in a low intensity/hard
spectrum regime in the LS and in a high intensity/soft spectrum regime
in the HS. The spectral modeling of the persistent emission confirms
this picture \cite{1916:bloser00apj}.  Assuming a source distance of
9.3 kpc, we derived a 2-20~keV luminosity of $1.4\times10^{37}$
\ergs~on March 13th (when \xb~is the brighest and in the HS) and of
$3.2\times10^{36}$~\ergs~on May 18th (when \xb~is the faintest and in
the LS).

\subsection{Variability in the $4\times10^{-3}-128$~Hz frequency
  range}

\subsubsection{Average power density spectra}

In order to better characterize the source, we looked for possible
correlations between the aperiodic variability and the spectral
behaviour.  We thus averaged together low frequency PDS of the LS on
one hand (Fig. \ref{summedpds} left), and PDS of the HS on the other
hand (Fig. \ref{summedpds} right), provided that they were obtained
from the same timing mode data.

\begin{figure}[!b]
\resizebox{\hsize}{!}{\includegraphics{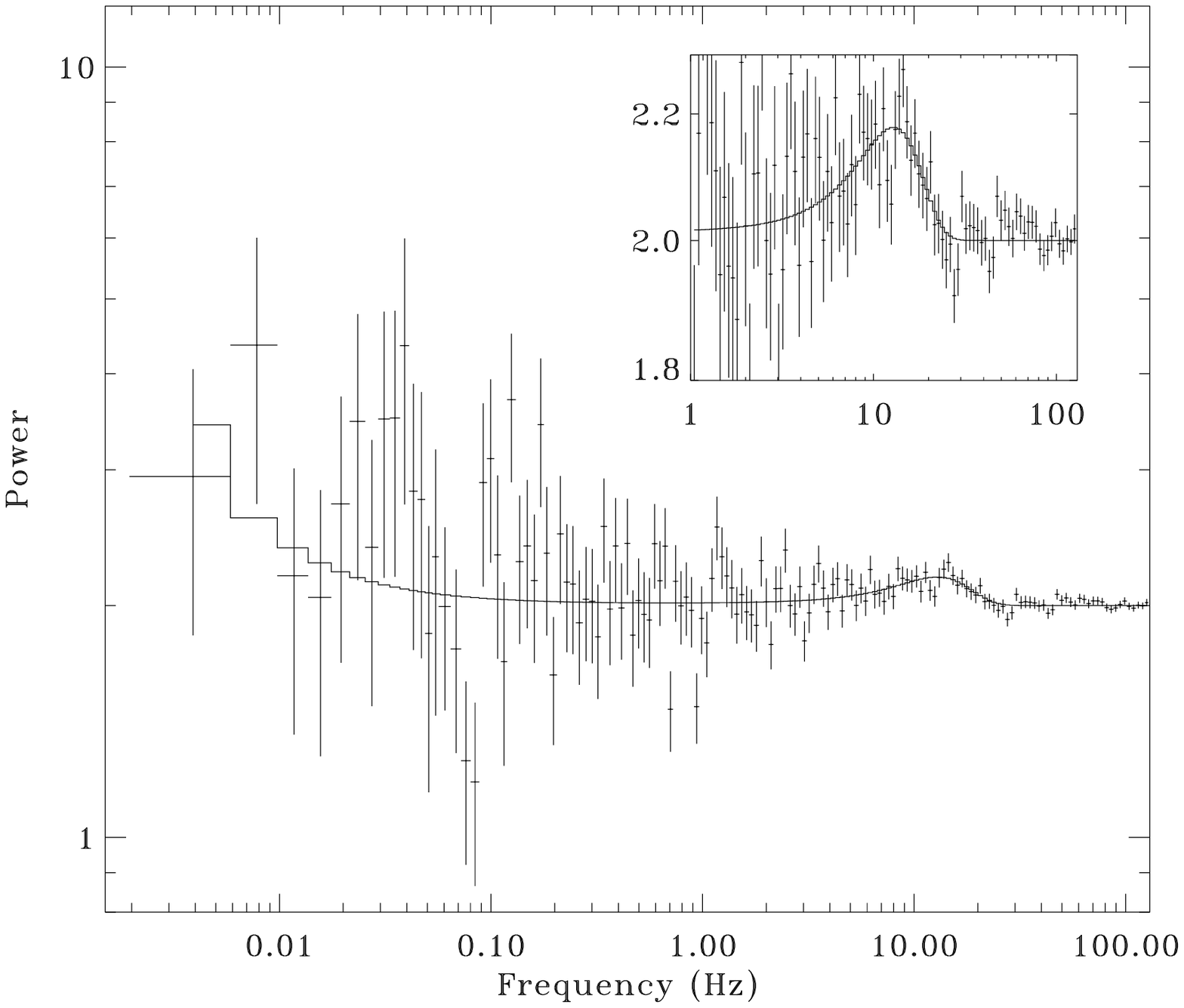}}\\%
\resizebox{\hsize}{!}{\includegraphics{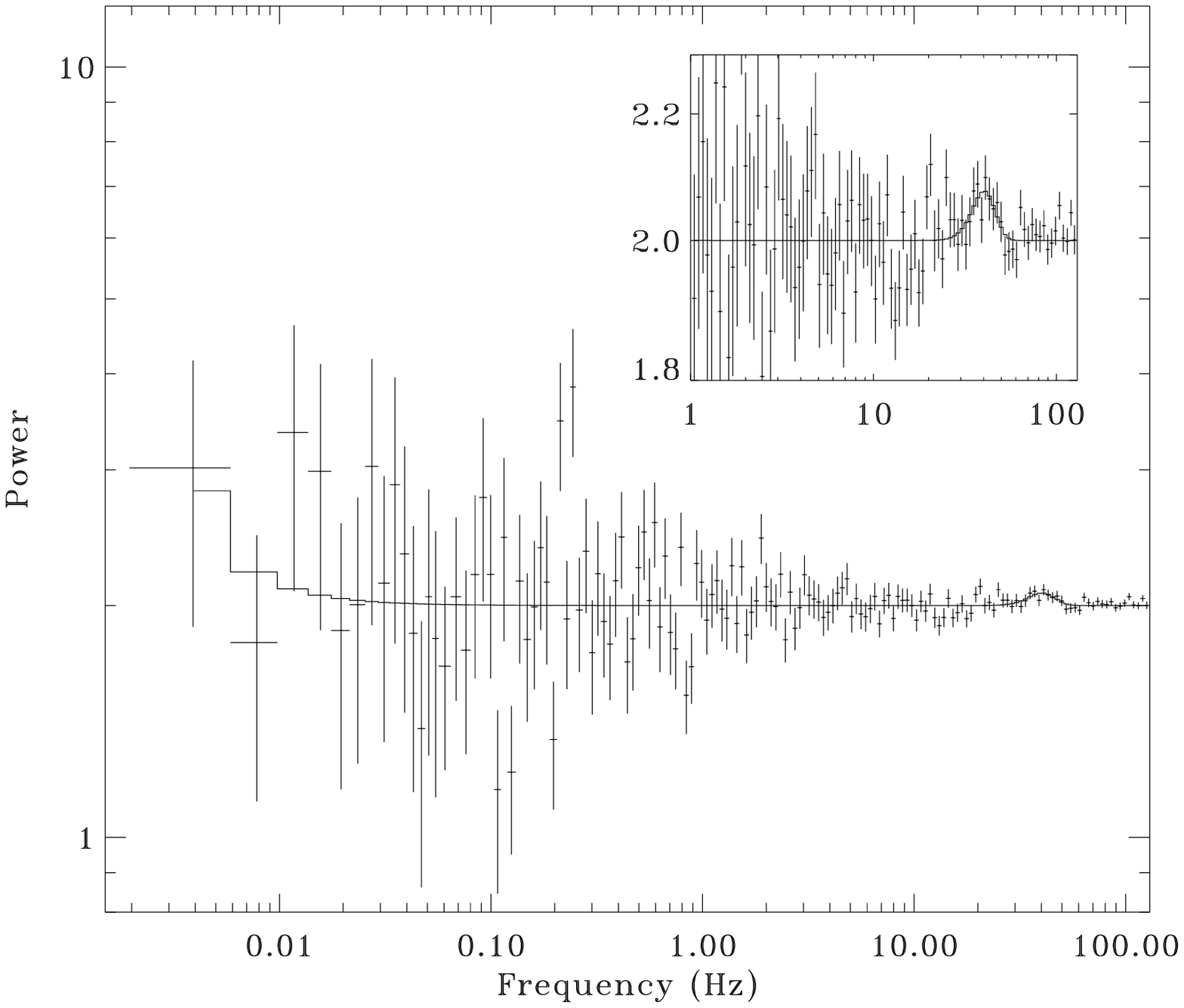}}
\caption{PDS showing LFQPOs in the persistent emission of two segments of
  observations on October 29th (top panel) and June 1st (bottom
  panel). The mean count rate is 35~\ctsspcu~for the first PDS and
  41~\ctsspcu~for the second (no background subtraction).  The
  exposure time of both PDS is 1800 seconds. The inset panel is a zoom
  of the PDS above 1~Hz. The solid lines are the best fits to the
  data.}
\label{low_freq_qpo}
\end{figure}

The average PDS of Fig. \ref{summedpds} (left) has an integrated RMS
over the entire $\sim4\times10^{-3}-128$~Hz frequency range of
$18.8\pm0.6$~\%. The PDS shows VLFN up to roughly 1~Hz and HFN
around 20~Hz. The model C+PL+CPL provides a better fit than a simple
PL (\chdof~$\sim$~166/146 and 324/149 respectively).  Looking at Fig.
\ref{summedpds} (right), we note an excess power around 0.2~Hz. We have
fitted that excess with a Gaussian and assessed its significance
through the standard F-test. We define P as the probability of
rejecting the hypothesis that the fit is better with the additional
Gaussian component. In the present case, we have P = 0.09, which means
that the QPO is significant at a level of $\sim$~91~\%. With the 0.2
Hz Gaussian added, the results of the best fit are given in Table
\ref{fit}.

We have searched for the same signal in different energy ranges. As an
example, in the 2-30~keV band, the significance of the 0.2~Hz signal
reaches 93~\%. The best fit parameters are given in Table \ref{fit}. In
addition to the state averaged PDS, we computed PDS for each observing
day. In some of these PDS, power excesses are visible around 0.2~Hz.
However, they are statisticaly marginal with significance levels less
than 90~\%.

In the HS, the source displays very low variability.
The total RMS in the 5-30~keV energy range is $5.3\pm0.2$~\%. At low
frequencies, the PDS is still dominated by VLFN up to roughly 0.1~Hz.
But above this frequency, the PDS is consistent with having no power.
No QPO nor HFN component are present in the PDS. The best fit
parameters are given in Table \ref{fit}.

\begin{figure}[!b]
\resizebox{\hsize}{!}{\includegraphics{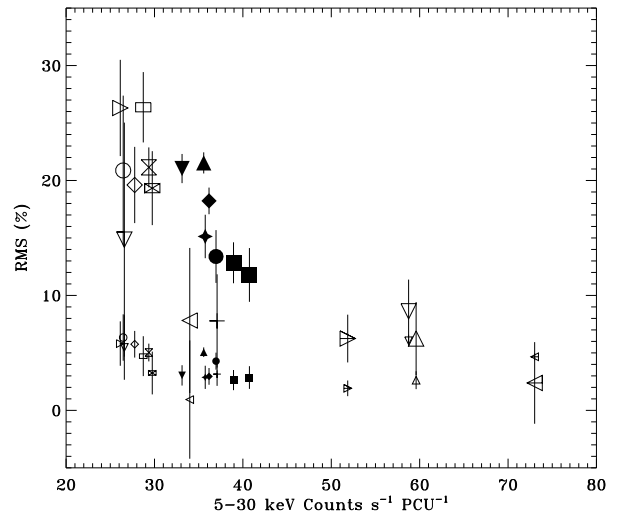}}
\caption{RMS in the $4\times10^{-3}-1$~Hz range (small symbols) and 
  in the 1-128~Hz range (big symbols) versus count rate (no background
  subtraction).}
\label{rms_rate}
\end{figure}

Looking at each individual PDS, we found that what we called
previously the HFN component in the state averaged PDS could be in fact
resolved into broad QPO-like features of varying centroid frequencies.
In the following, we call these features LFQPOs.  Fig.
\ref{low_freq_qpo} shows two examples of daily-averaged PDS where such
a LFQPO is clearly detected.  Detailed properties of the LFQPOs are
given in the next section.  To conclude, \xb~displays aperiodic
variability below $\sim$~1~Hz (the VLFN) and quasi-periodic
variability or no variability above 1~Hz.

Fig. \ref{rms_rate} displays the RMS as a function of the count rate for
each observation. The RMS is given in the $3\times10^{-3}-1$~Hz
frequency range where the VLFN dominates, and between 1-100~Hz where
the LFQPO is detected.  The RMS of the VLFN component ranges between
$\sim$~3 and 6~\%, whereas the RMS above 1~Hz ranges between $\sim$~5
and 25~\%.  In the LS, the RMS is high and decreases as the average
count rate per observation increases. On the other hand, in the HS,
the RMS, after reaching a low threshold, does not significantly vary
as the count rate increases.  Thus, the timing behaviour of \xb~is
different in its two states.

\subsubsection{Low frequency quasi-periodic oscillations}
\label{sec:lfqpos}

To further investigate the properties of the LFQPOs detected between a
few~Hz and 100~Hz in \xb~(see Fig. \ref{low_freq_qpo}), and given
their relative weakness in the persistent emission, we searched for
them in the persistent and dipping emission combined rather than in
the persistent emission alone. Although we note that we may be looking
at different signals when combining the persistent and dipping
emissions, we aimed to get the best signal to noise ratio.  The study
of the LFQPOs is carried out in the 5-30~keV energy range in order to
make easier the comparison with the HFQPOs detected in that energy
range with the best signal to noise ratio (see next section).  The
dipping is large but not complete in that energy range.  So, adding
the dipping parts, which represent roughly 40~\% of the data set,
actually increases the signal to noise ratio.  This is not negligible
because \xb~is relatively faint in the PCA (2-20~keV background
subracted count rate around 35~\ctsspcu, to be compared with
$\sim$~30000~\ctsspcu~for Sco X-1) and observed with only 3 PCUs for
more than the half of our observation (see Table \ref{exposure}).
Furthermore, the ingress or egress of a dip lasts more than 1 second
(typically 40 s).  So, above 1~Hz where the LFQPOs are detected, the
PDS of the persistent and dipping emission is not contaminated by the
dip phenomenon itself.  PDS were computed in the range
$\sim4\times10^{-3}-128$~Hz the same way as for the persistent
emission alone.  We averaged together PDS of the same observation or
the same segment.

\begin{figure}
\resizebox{\hsize}{!}{\includegraphics{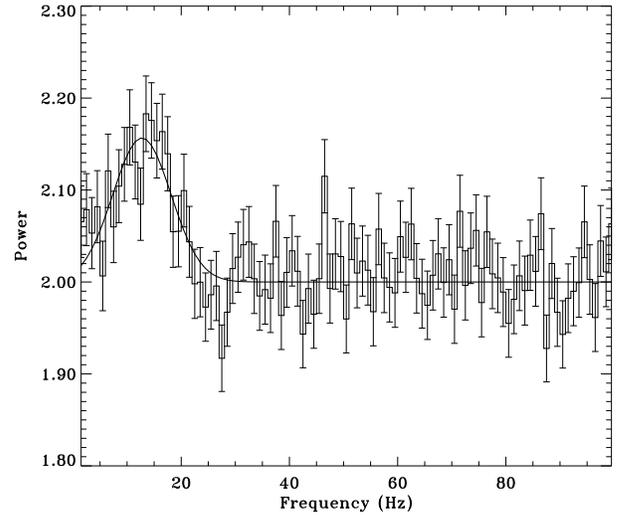}}
\caption{PDS showing the LFQPO in a segment of the October 29th
  observation (persistent and dipping emissions combined). The solid
  line is the best fit of the QPO with a Gaussian.}
\label{qpo2910}
\end{figure}

The LFQPOs frequency, FWHM, RMS and significance level are listed in
Table \ref{tabqpo}.  Their frequency ranges between 5 and 80~Hz. Their
FWHM is around 16~Hz but can vary from 0.5 to 45~Hz, so that their
coherence (Q=$\nu$/FWHM) changes roughly from 0.5 to 26 with a mean
value of 4. The designation of QPO is then sometimes improper since
this is usually used for signals with Q values greater than 2
\cite{vdk95xrb}.  One of the LFQPOs is shown in Fig.  \ref{qpo2910}.
In the second segment of September 6th, we found an excess in the PDS
between 1 and $\sim$~50~Hz ($\sim6.3~\sigma$). A visual inspection of
the PDS indicates that this excess is flat, yet above the noise level.
For this reason, we were unable to fit it with a Gaussian, and
therefore we do not consider this excess as a LFQPO, as defined above.
Interestingly enough, however we note that the centroid of the excess
($\sim$~30~Hz) would fall on the correlation between LFQPO frequency
and the upper HFQPO frequency, as shown in Fig. \ref{bf_vs_hf} (see
section \ref{sec:lfhf}).
\begin{table*}
\begin{center}
\scriptsize
\renewcommand{\arraystretch}{1.3}
\newcolumntype{.}{D{.}{.}{4}}
\newcolumntype{b}{D{.}{.}{1}}
\begin{tabular}{c...b...bc}

\hline
Obs & \multicolumn{4}{c}{ LFQPO}& \multicolumn{5}{c}{ HFQPO}\\
& \multicolumn{1}{c}{$\nu$} &\multicolumn{1}{c}{FWHM} &
\multicolumn{1}{c}{RMS } &
\multicolumn{1}{c}{$\sigma$ } & \multicolumn{1}{c}{$\nu$} &
\multicolumn{1}{c}{FWHM} &\multicolumn{1}{c}{RMS}  &
\multicolumn{1}{c}{$\sigma$} &
\multicolumn{1}{c}{$\Delta\nu$} \\
& \multicolumn{1}{c}{(Hz)} &\multicolumn{1}{c}{ (Hz)} &
\multicolumn{1}{c}{ (\%)} & & \multicolumn{1}{c}{ (Hz)} &
\multicolumn{1}{c}{ (Hz)} &\multicolumn{1}{c}{ (\%)} & &
\multicolumn{1}{c}{ (Hz)} \\
\hline
\hline 
02/10 *&  & & 4 &&&&7&&\\
03/13 *&&&3 &&&&6&&\\
05/05 1& 72.3^{+  6.5\; a}_{- 6.3} &  22.1^{+ 12.7}_{- 11.1 } &
10.8\pm 1.5 &3.6 & & &13 & \\
05/05 2  &&&9&&1263.9^{+1.8\; a}_{-1.7} & 6.8^{+3.8}_{-3.1}& 8.1\pm1.1&3.7& \\ 
05/06 1&&&10&&879.9^{+2.9}_{-2.7} & 12.8^{+5.4}_{-5.2} &  9.9\pm 1.1
& 4.5&\\
05/14 1 &&& 18 && 946.7^{+ 16.1}_{- 16.1} &  77.2 ^{+ 28.8}_{-26.1} & 32.9 \pm 3.4 &4.8\\
05/14 2 &   29.2 ^{+  0.6\; a}_{-  0.6} &   1.6 ^{+
  1.5}_{-  1.0} &  8.3 \pm  1.4 & 3.0 &960.3^{+0.8\; a}_{-0.9} &2.8^{+2.2}_{-1.5} & 10.2\pm1.5&3.4& \\
05/14 3 &  && 17&& 839.5 ^{+ 26.0\; a}_{- 26.6} &  93.9
^{+ 46.5}_{- 33.5} & 32.0 \pm  4.4  & 3.6&\\
05/15 1 &    & 
  &  13& & 1055.3 ^{+ 17.1}_{- 17.9} &
80.4 ^{+ 35.8}_{- 29.9} & 19.4 \pm  2.1 &4.6& \multirow{2}{0.7in}{348.4$^{+ 28.1 }_{- 27.4}$}\\
&  & &  & &706.9 ^{+ 22.4\; a}_{- 20.9} &  73.4 ^{+39.5}_{- 36.5} & 16.4 \pm  2.4 & 3.4& \\
05/16 *&   31.1 ^{+ 13.6\; a}_{- 15.7} &  45.2 ^{+
  29.0}_{- 19.4} & 11.3 \pm  1.7 & 3.3& 1006.4 ^{+ 16.2}_{- 16.6} &
79.8 ^{+ 37.2}_{- 29.9} & 16.4 \pm  1.7 &4.8&\multirow{2}{0.7in}{349.5$^{+  22.9 }_{- 24.7}$}\\
 & &    && & 656.9 ^{+ 16.2\; a}_{- 18.3} &  53.7 ^{+ 32.5}_{- 27.1} & 12.0 \pm  1.9 & 3.2\\
05/17 *&   &&8&& 942.0 ^{+ 27.0}_{- 26.1} & 131.8 ^{+ 59.3}_{-
  44.3} & 21.7 \pm  2.2 & 4.9&\\
05/18 1 & && 16&&  577.2 ^{+ 32.3}_{- 31.6} & 146.5 ^{+ 64.6}_{-
  53.7} & 36.3 \pm  3.9 &4.7\\
05/18 2&  20.3^{+  4.3\; a}_{- 4.6} &  15.2^{+8.6 }_{-6.8} &  18.6\pm
2.7&3.4  &&& 23 &&\\
05/19 1 &   19.1 ^{+  2.8}_{-  2.8} &  12.2 ^{+  7.1}_{-
  5.9} & 19.8 \pm  2.2&4.5 & & &24 && \\
05/20 * &17.5^{+2.9 }_{- 3.3}&12.4^{+6.32}_{-5.0} & 13.8\pm1.6 &4.3& & & 20& &\\
05/21 1 &    8.0 ^{+  0.4}_{-  0.4} &   2.0 ^{+  0.7}_{-
  0.6} & 12.5 \pm  1.3&4.8 & & &21 && \\
05/21 2 &   12.6 ^{+  0.1\; a}_{-  0.1} &   0.5 ^{+
  0.2}_{-  0.2} &  8.1 \pm  1.1 &3.7& & & 25&& \\
05/22 *& 14.0 ^{+3.2}_{- 3.4} &  18.6^{+9.4 }_{-5.7} & 16.3\pm1.3 &6.3& &&17&&\\
05/23 1 &   34.4 ^{+  4.0\; a}_{-  3.0} &  11.8 ^{+  8.1}_{-
  4.7} &  8.5 \pm  1.1&3.9 &  818.9 ^{+ 11.4}_{- 10.9} &  51.2 ^{+
  23.1}_{- 18.4} & 13.9 \pm  1.5 & 4.6&\\
05/23 2 &   31.8 ^{+  2.2}_{-  2.4} &  12.6 ^{+
  6.3}_{-  3.9} & 10.6 \pm  1.0 & 5.3& 845.5 ^{+ 15.9}_{- 15.9} &  78.4 ^{+ 36.2}_{- 25.5} & 15.9 \pm  1.6 & 5.0&\\
06/01 1 &   36.6 ^{+  6.2}_{-  7.0} &  35.0 ^{+
  20.0}_{- 12.2} & 12.1 \pm  1.0 & 6.0& 935.2 ^{+ 13.3}_{- 13.3} &  89.0
^{+ 30.1}_{- 25.2} & 16.6 \pm  1.2 & 6.9&\\
06/01 2 & 35.2 ^{+0.7\; a}_{-0.7} & 2.2^{+1.5}_{-1.2}&4.9\pm0.8 &3.1& 911.3 ^{+ 16.8}_{- 16.4} & 103.2 ^{+ 43.8}_{-
  28.3} & 18.3 \pm  1.5 &6.1&\multirow{2}{0.7in}{352.9$^{+17.3}_{-17.0}$} \\
&  &  & & &558.3 ^{+  4.3}_{-  4.6} &  18.4 ^{+  8.8}_{-
  6.9} &  9.8 \pm  1.2 &4.1&  \\
06/01 3 &&&7&&953.1 ^{+ 13.9}_{- 13.9} &  71.7 ^{+24.8}_{- 20.4} & 14.5 \pm  1.4 &5.2&\\

07/15 *&   15.3 ^{+  2.1}_{-  2.3} &  26.1 ^{+  5.8}_{-
  4.8} & 15.4 \pm  0.6 &12.8&  710.0 ^{+ 31.5}_{- 30.7} & 149.1 ^{+
  73.9}_{- 48.7} & 17.8 \pm  1.4 & 6.4&\\
08/16 1 &12.6 ^{+  2.2}_{-  2.4} &  11.3 ^{+5.0}_{-  3.9} & 19.5 \pm  2.0&4.9& 513.6^{+0.2}_{-0.2} &1.2^{+0.6}_{-0.4}&12.4\pm1.3 &4.8&\multirow{2}{0.7in}{290.0$^{+4.7}_{-4.2}$} \\
&&&&&223.6^{+4.7\;a}_{-4.2}&16.2^{+8.6}_{-5.9}&18.3\pm2.5& 3.7&\\
08/16 2 &   11.8^{+  1.4}_{-  1.4} &  11.7^{+3.8}_{-2.9} & 15.8 \pm  0.9&8.8&  538.1^{+ 15.3\; a}_{- 14.8} &  55.4^{+ 27.1}_{- 24.2} & 15.5 \pm  2.1 &3.7&\\
08/16 3 &    8.7 ^{+  0.9}_{-  1.0} &  12.2 ^{+  2.2}_{-  1.9} & 19.6 \pm  0.7 &14.0& & &15 && \\
08/16 4 &   12.9 ^{+  1.2}_{-  1.2} &   6.9 ^{+  3.7}_{-
  2.8} & 12.3 \pm  1.0 &6.1& & & 18 && \\
09/06 1 &   28.2 ^{+  9.0}_{-  9.9} &  44.5 ^{+
  23.9}_{- 16.1} & 15.3 \pm  1.5 &5.1& 844.1 ^{+ 11.5}_{- 10.8} &  73.0 ^{+ 22.4}_{- 19.5} & 20.2 \pm  1.5 & 6.7&\\
09/06 2 & 30^b & 20 & 17& 6.3
&808.4^{+9.5}_{-10.5} &39.6^{+21.8}_{-17.1} & 14.4 \pm1.8& 4.0\\
09/06 3 &   34.0 ^{+  5.1}_{-  5.2} &  24.3 ^{+
  12.8}_{-  8.9} & 12.5 \pm  1.3 &4.8&  888.6 ^{+ 32.4}_{- 30.2} &
172.8 ^{+ 68.4}_{- 66.7} & 22.5 \pm  2.0 &5.6& \\
09/06 4 &   45.7 ^{+  2.8\; a}_{-  3.1} &  10.0 ^{+
  6.3}_{-  4.6} &  8.2 \pm  1.2 &3.4&  923.8 ^{+ 23.2}_{- 22.7} &
125.4 ^{+ 56.6}_{- 45.0} & 19.5 \pm  1.8 &5.4& \multirow{2}{0.7in}{340.5$^{+ 24.2}_{- 23.6}$}\\
 &  & & & & 583.2 ^{+  6.8\; a}_{-  6.5} &  21.5 ^{+ 20.6}_{-
   10.8} &  9.7 \pm  1.4 &3.5& \\
10/29 1 &   12.9 ^{+  0.7}_{-  0.8} &  10.2 ^{+  1.5}_{-  1.3} & 17.1
\pm  0.6  &14.2& 610.9^{+24.4\; a}_{-25.9} &  88.4^{+ 48.9}_{- 38.9} &
14.8 \pm  2.1 &3.5& \\
10/29 2 & 8.5 ^{+2.0}_{-2.0} & 17.5^{+5.6}_{-3.8} & 16.4\pm0.8 &10.2&
573.0^{+1.3\;a}_{-1.4} &4.6^{+2.8}_{-2.3} & 7.6\pm1.1 &3.5& \\
10/29 3 &   20.6 ^{+  2.7}_{-  2.8} &  25.0 ^{+
  7.9}_{-  6.3} & 16.8 \pm  0.8 &10.5&  610.2 ^{+  4.1}_{-  4.2} &  20.7 ^{+  8.5}_{-  6.4} & 11.9 \pm  1.2 &5.0& \\
\hline
\end{tabular}
\end{center}
\caption{Properties of the QPOs detected above 3~$\sigma$ (see text) in \xb~in
  the 5-30~keV energy range. We list the observation date (month/day) 
  followed by the segment number or by an asterisk if all segments
  have been summed, the frequency, FWHM, RMS and significance  of the
  LFQPOs (1-100~Hz) and  HFQPOs (100-1400~Hz). The last column is the
  separation between the twin HFQPOs. All the  values are inferred from 
  Gaussians used to fit the QPOs. The reduced $\chi^2$ are close to 1.
  The error bars are given for the 90~\% confidence level
  ($\Delta\chi^2$=2.7 for the variation of a single parameter). When
  no signal is present above our confidence threshold, we give a 3~$\sigma$ upper limit on the RMS of a signal of FWHM 10~Hz for  the
  LFQPOs and 50~Hz for the HFQPOs. The significances quoted here are
  for a single trial only. A more detailed assessment of the
  significance of the detections taking into account the number of
  trials may be found in the text (see section
  \ref{sec:search_technique}). $^{a}$ flags detections at a
  significance level between 3 and 4~$\sigma$. The other signals have
  significance level greater than 4~$\sigma$. $^{b}$ The excess could
  not be fitted as a Gaussian (see text). Approximated values of the
  frequency, FWHM and RMS are given.} 
\label{tabqpo}
\end{table*}

\begin{figure*}
\resizebox{\hsize}{!}{\includegraphics{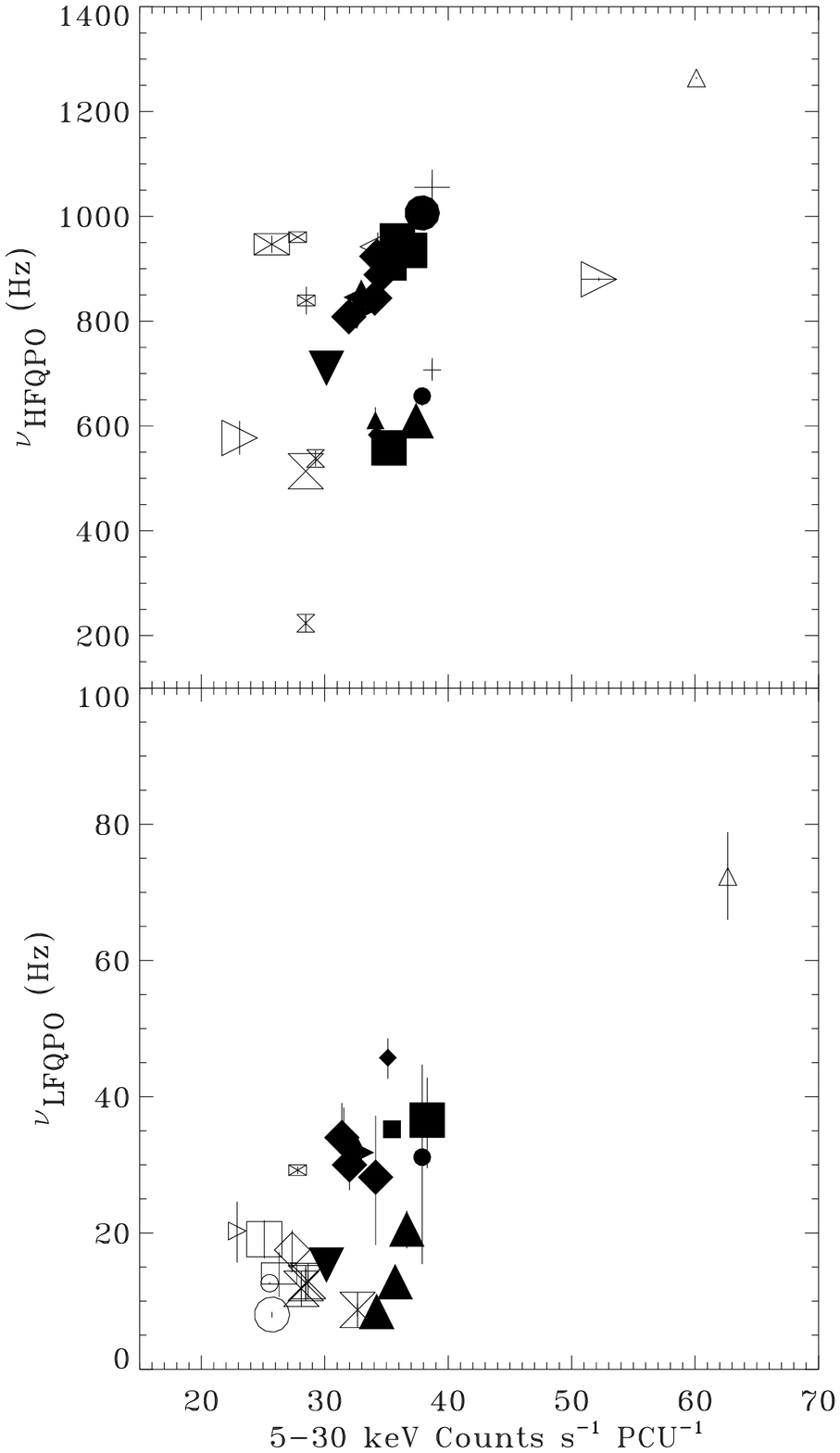}{\includegraphics{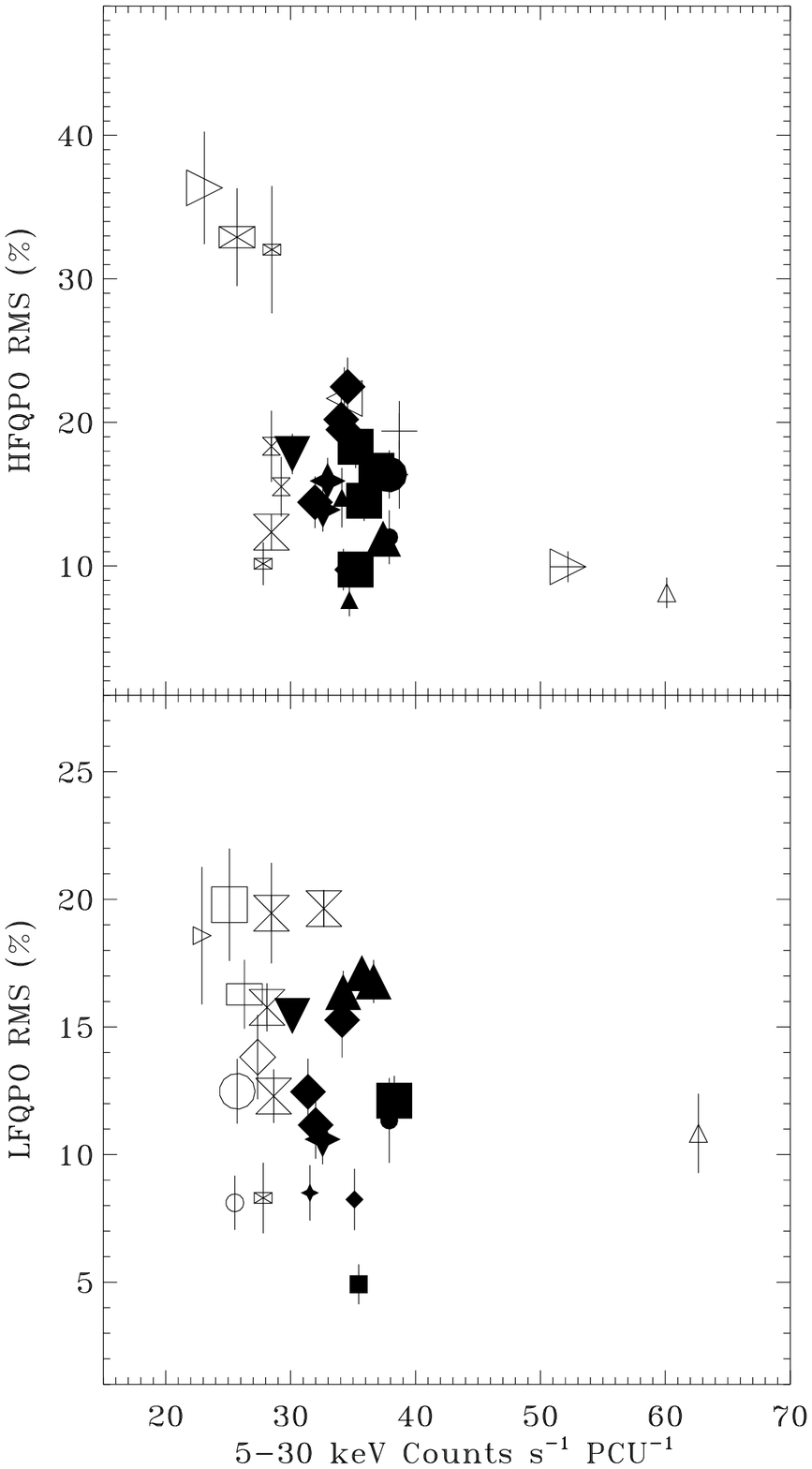}}}
\caption{Left: LFQPO (bottom panel) and HFQPO (top panel) frequency as a
  function of the 5-30~keV count rate (no background subtraction).
  Right: LFQPO (bottom panel) and HFQPO (top panel) RMS as a function
  of the 5-30~keV count rate.  Small symbols are used for detections
  between 3 and 4~$\sigma$, big symbols for detections above 4~$\sigma$.}
\label{nu_hf_bf_vs_cts}
\end{figure*}

\begin{figure*}
\resizebox{\hsize}{!}{\includegraphics{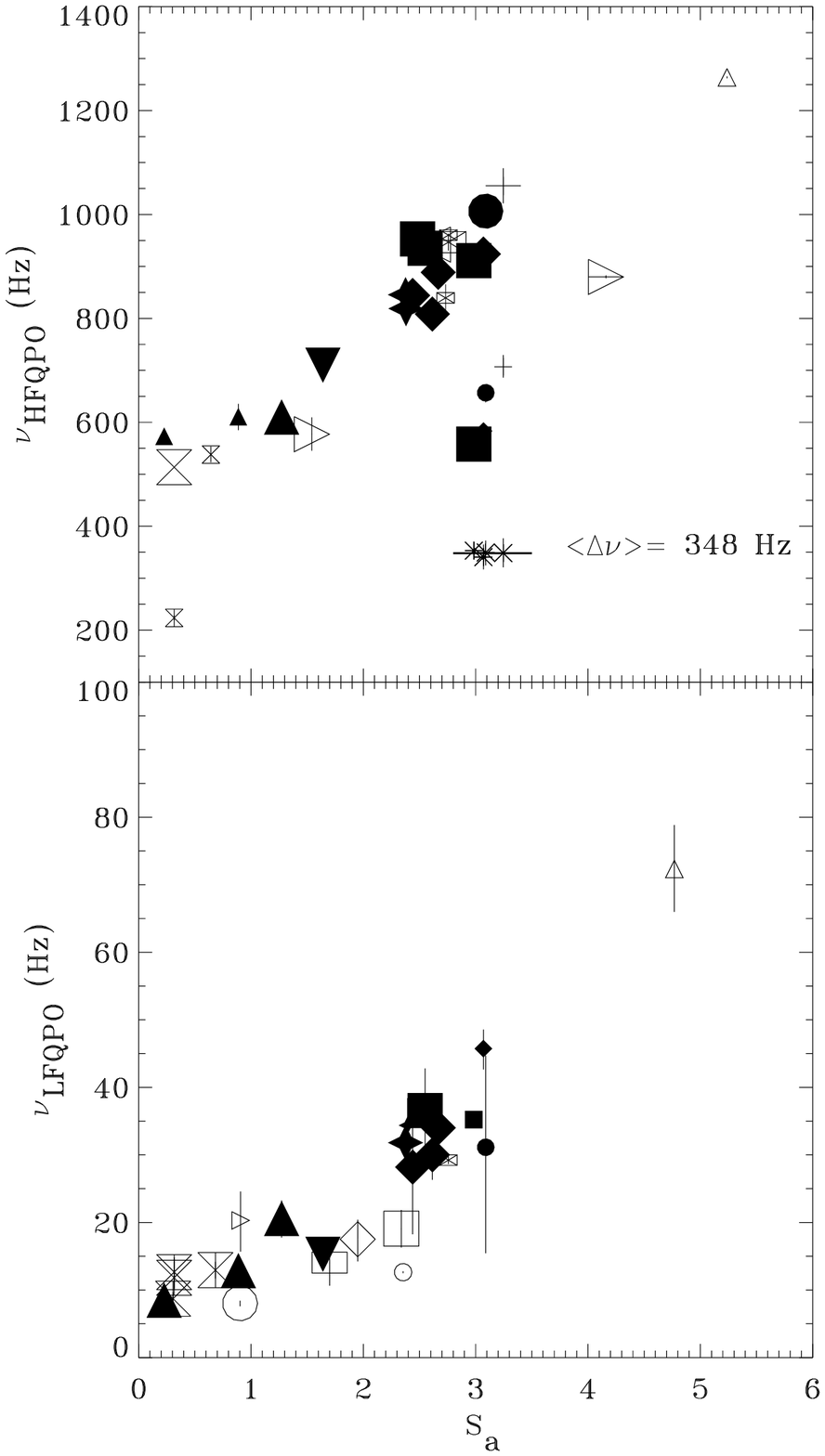}\includegraphics{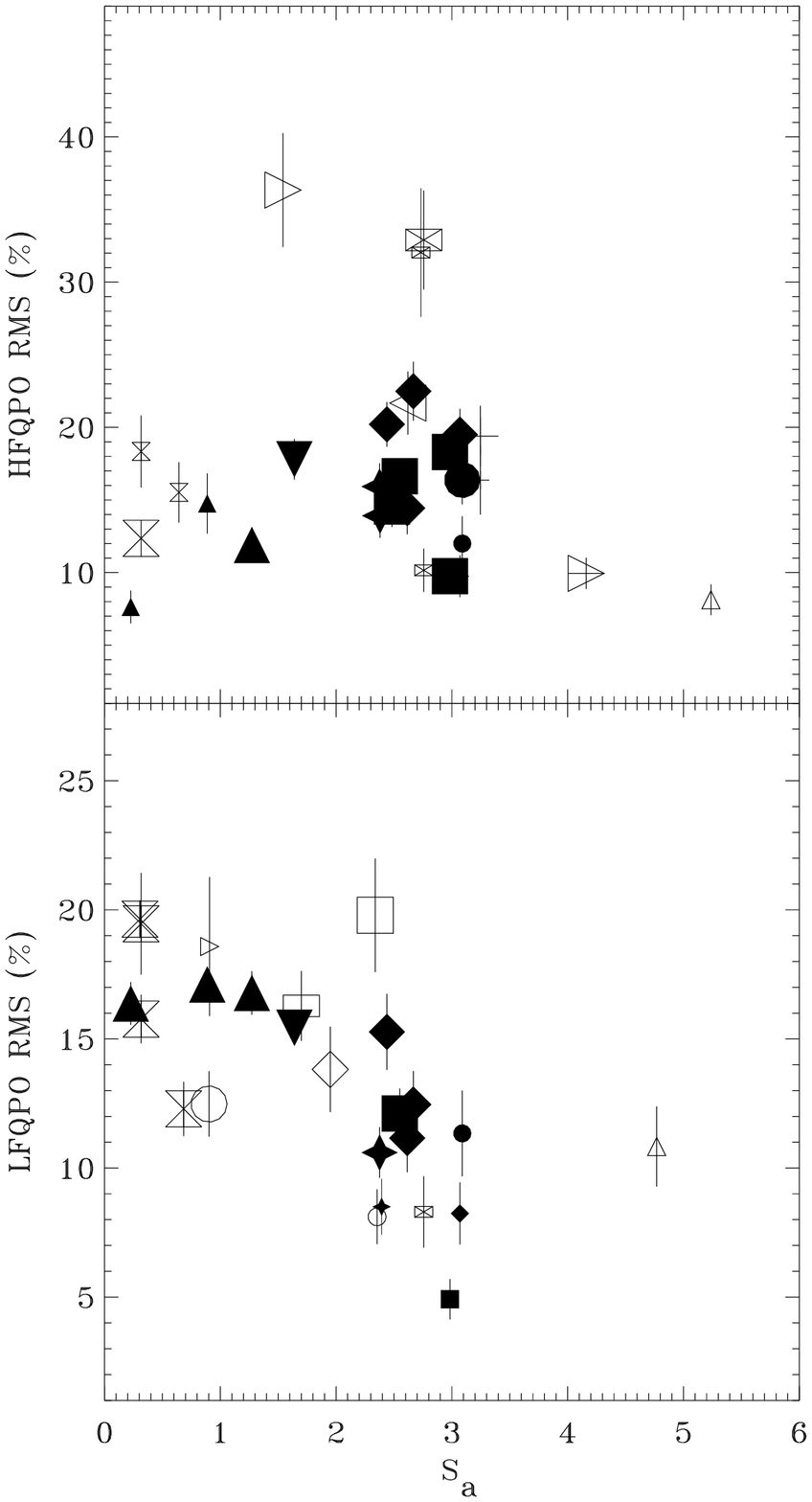}}
\caption{Left: LFQPO (bottom panel) and HFQPO (top panel) frequency as a
  function of the position \sa~on the color-color diagram. The
  asterisks represent the frequency separation of the twin HFQPOs. The
  solid line is the average separation.  Right: LFQPO (bottom panel)
  and HFQPO (top panel) RMS versus \sa. Small symbols are used for
  detections between 3 and 4~$\sigma$, big symbols for detections
  above 4~$\sigma$.}
\label{nu_hf_bf_vs_sa}
\end{figure*}

We looked for correlations between LFQPOs and spectral states.  The
bottom-left panel of Fig. \ref{nu_hf_bf_vs_cts} shows the LFQPO
frequency versus the 5-30~keV count rate. All but one of the LFQPOs
are detected in observations of the LS. For \xb, the LFQPO frequency
seems to roughly correlate positively with the intensity, but the
scatter is substantial.

The bottom-right panel of Fig. \ref{nu_hf_bf_vs_cts} shows the LFQPO
RMS as a function of count rate.  The RMS ranges from 5 to 20~\%.  For
the HS, we have derived upper limits ranging between 3 and 10~\% (see
Table \ref{tabqpo}). No simple relation seems to connect the two
quantities. 

Recent works have shown that some timing features of similar systems
were better correlated with the position on the color-color diagram
than with the count rate \cite{1608:mendez99apjl,1728:mendez99apjl}.
We thus examined the QPO properties as a function of the position on
the color-color diagram.

Color-color diagrams are sensitive to changes of a few percent in flux
ratios between energy bands.  Observations allowed to show that during
these subtle changes, the X-ray intensity was not a measure of the
accretion rate since both positive and negative correlations between
the two quantities could occur \cite{hasinger89aa,vdk94apjs}.  The
position on the color-color diagram is a better indicator of the
spectral state and hence the accretion rate than the count
rate.

The direction of evolution of the accretion rate on the color-color
diagrams was inferred from X-ray luminosities and confirmed by
multiwavelength and bursts observations
\cite{cygx2:hasinger90aa,1636:vdk90apjl}. In the atoll sources, the
accretion rate is believed to increase from the island to the lower
and then to upper banana states. For our analysis, we refer to the
color-color diagram shown in Fig.  \ref{color_color}. The spline shows
the approximated track followed by \xb~during the observations.  The
direction of evolution of the accretion rate along the track is
derived from the X-ray luminosity which globaly increases from the
extremity located at the top of the diagram to the extremity located
at the bottom (see Bloser et al. \cite*{1916:bloser00apj} for details
about spectral fitting).

Following Méndez et al.  \cite*{1608:mendez99apjl}, we called \sa~the
parametrization along this spline. \sa~is a measure of the position of
the source on the diagram. \sa~was choosen to increase from top to
bottom so that it is representative of the inferred evolution of the
accretion rate.  We determined the value of \sa~for each segment of
observation by projecting the corresponding point of the color-color
diagram (inset panel of Fig. \ref{color_color}, where one point
corresponds to one segment) on the spline. This value of \sa~could
then also be attributed to the LFQPO detected in the given segment.
For LFQPOs detected in PDS averaged from several segments of the same
observation day, we used the mean of the \sa~values found for each
segment.  The bottom-left panel of Fig.  \ref{nu_hf_bf_vs_sa} shows
the LFQPO frequency as a function of \sa.  A positive correlation is
now clearly visible. In the bottom-right panel of Fig.
\ref{nu_hf_bf_vs_sa}, we show that the LFQPO RMS anticorrelates with
\sa. In both plots, the scatter is now much smaller than in the
previous plots where the count rate was involved.  Note that we have
also studied the evolution of the LFQPOs as a function of the hard
color (10-30~keV/5-10~keV). We concluded that the hard color was a
worse indicator of the timing behaviour than \sa: the evolution of the
soft color had to be taken into account.

\subsection{Discovery of high frequency quasi-periodic oscillations}

\begin{figure}[!t]
\resizebox{\hsize}{!}{\includegraphics{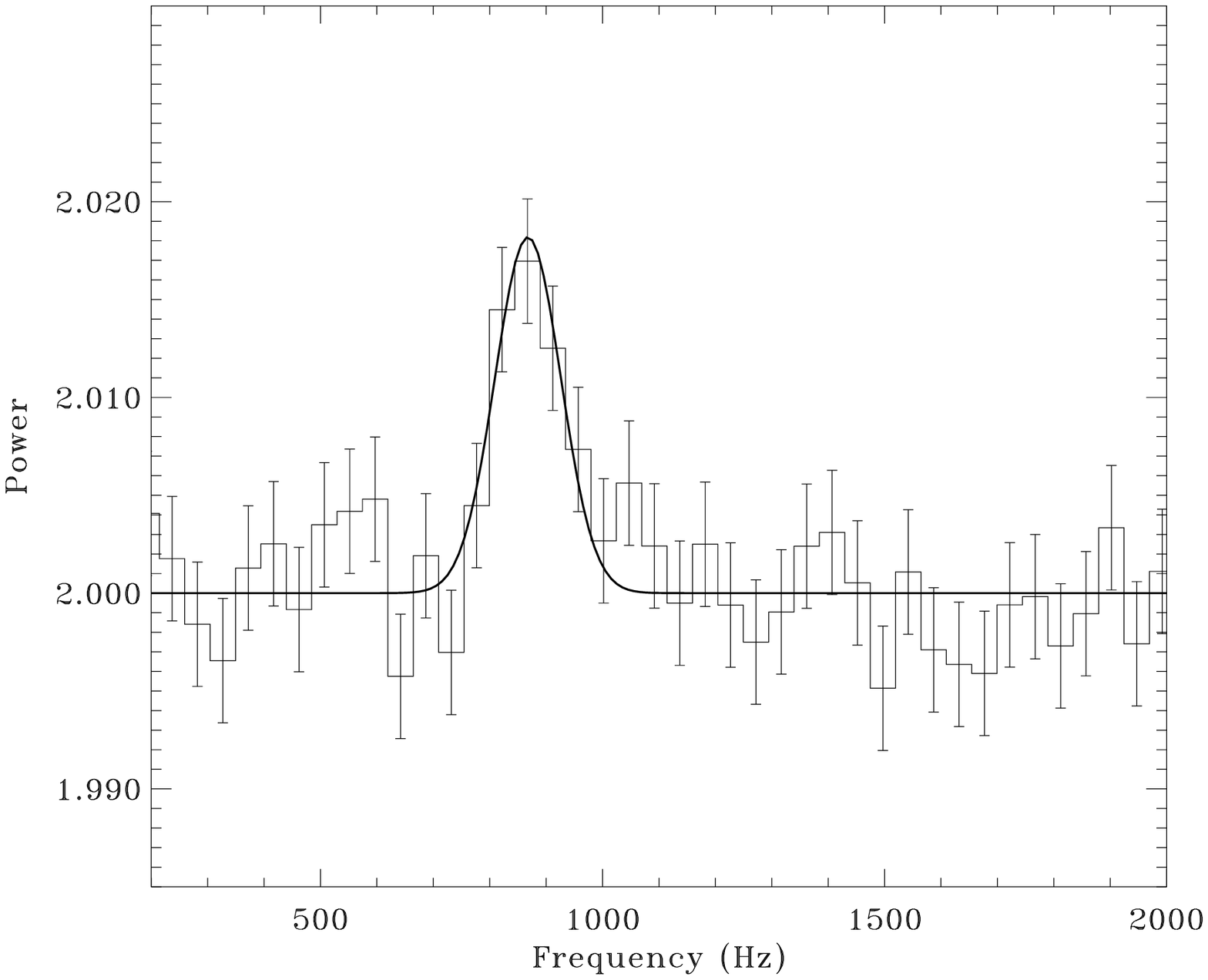}}
\caption{PDS for the September 6th observation. The HFQPO is fitted
  with a Gaussian.} 
\label{strong_qpos}
\end{figure}

\begin{figure}[!t]
\resizebox{\hsize}{!}{\includegraphics{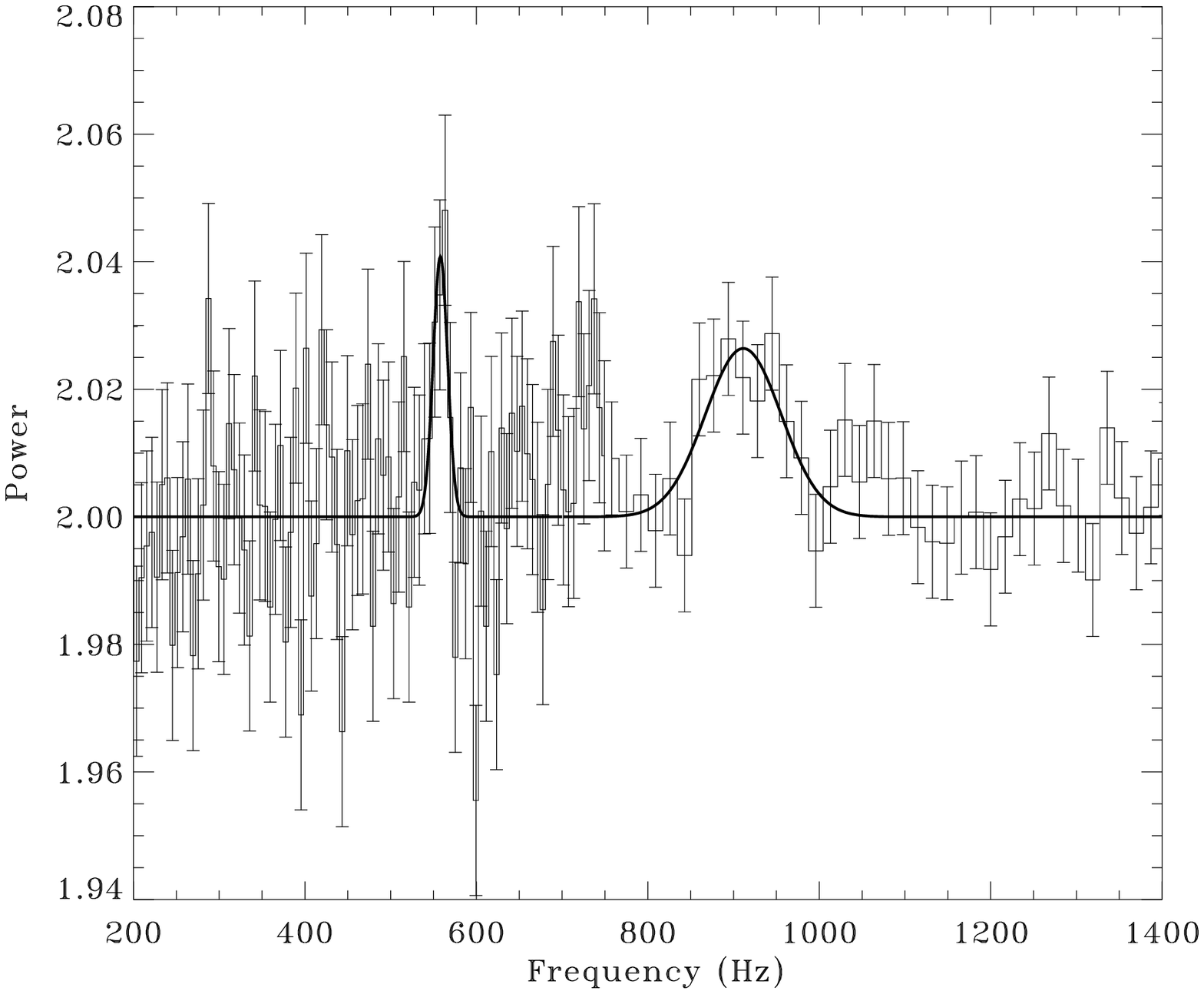}}
\caption{The twin HFQPOs fitted with Gaussians detected in a segment
  of the June 1st observation. For display purposes, two different
  frequency binnings have been used.}
\label{twin}
\end{figure}

Now, we report on the discovery of HFQPOs from \xb.  The analysis has
been carried out in the 5 to 30~keV energy range because in other
sources, it is known that HFQPOs are more easily detected in this
energy band, as in, e.g.  KS1731-260 \cite{1731:wijnands97apjl}. The
persistent and dipping emission were combined to increase the signal
to noise ratio (see section \ref{sec:lfqpos}).

The results are reported in Table \ref{tabqpo} (either the
observations or the segments). The HFQPOs frequency ranges between
$\sim$~200 and 1300~Hz. Their FWHM is around 70~Hz and coherence
around 50. Their RMS amplitude is high ($\sim$~17~\%).
One of the strongest HFQPO signals is shown in Fig.
\ref{strong_qpos}.

We detect twin HFQPOs in 5 segments or observations (Table
\ref{tabqpo}).  For four of them, the frequency separation is
consistent with being constant. The mean separation is $348\pm12$
Hz. For the fifth, the separation is $290\pm5$~Hz, thus inconsistent
with the previous value. We note that this pair of QPOs occur at 514
and 224~Hz whereas the four other twin peaks occur above 550~Hz.  The
twin HFQPOs detected in a segment of the June 1st observation are
shown in Fig. \ref{twin}.

HFQPOs were not detected in observations with the largest count rates
(February 2nd, March 13rd) with upper limits of $\sim$~6~\% on the RMS
(Table \ref{tabqpo}).  HFQPOs were not detected either in the lowest
count rate regime (May 18th to May 22nd); this might be due to a lack
of sensitivity. Indeed, only 3 PCUs were working during these
observations (see Table \ref{exposure}) and the upper limits on the
RMS are not constraining ($\sim$~20~\%, see Table \ref{tabqpo}).

The top-left panel of Fig. \ref{nu_hf_bf_vs_cts} shows the HFQPO
frequency as a function of count rate.  In this diagram, we observe a
branch between 700 and 1100~Hz where the frequency strictly correlates
with the count rate. A second parallel branch seems to be drawn at
lower frequencies.  However, some of the HFQPOs of that branch do not
have a simultaneous twin peak in the upper branch.  In the lowest
count rate regime, the frequency does not follow anymore a simple
relation with the count rate.

The top-left panel of Fig. \ref{nu_hf_bf_vs_sa} shows the HFQPO
frequency as a function of \sa.  The frequency versus \sa~relation
appears much simpler than the frequency versus count rate relation. We
can now easily identify the upper and lower HFQPOs. The frequency of
the upper HFQPO is now well correlated with \sa~within its full range.
The four twin peaks above 550~Hz are detected within a narrow range of
\sa~around 3.  Their frequency separation is shown with an asterisk on
Fig.  \ref{nu_hf_bf_vs_sa}.  The pair of QPOs at 224~Hz and 514~Hz
appears at the lowest \sa~and hence lowest inferred accretion rate.
Concerning the HFQPO RMS, we show that, surprisingly, it is better
correlated (negatively) with the count rate than with \sa~(top-right
panel of Fig.  \ref{nu_hf_bf_vs_cts} and \ref{nu_hf_bf_vs_sa}
respectively).

\subsection{Correlation between low  and high frequency quasi-periodic
  oscillations}

\begin{figure}[!t]
\resizebox{\hsize}{!}{\includegraphics{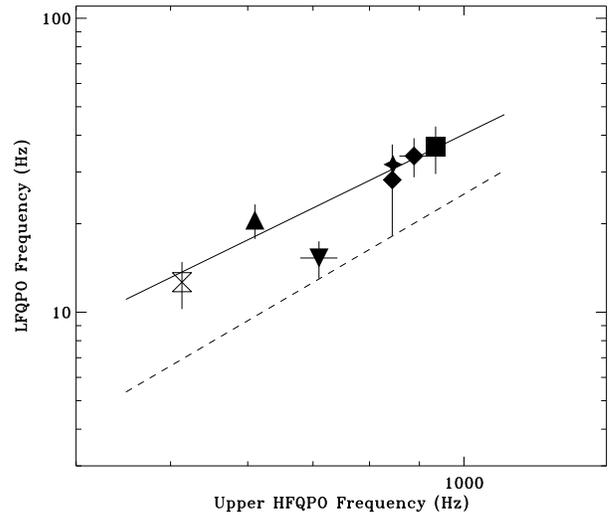}}
\caption{LFQPO frequency as a function of the upper HFQPO
  frequency. We report signals detected simultaneously above
  4~$\sigma$.  The solid line is the powerlaw of index 1.6 fitting the
  data. The dashed line is the precession frequency (Lense-Thirring
  and classical contributions (Stella and Vietri 1998)) assuming
  standard parameters for the neutron star (see text) and a spin
  frequency of 348~Hz.}
\label{bf_vs_hf}
\end{figure}

Fig. \ref{nu_hf_bf_vs_sa} (left panel) shows the parallelism between
HF and LF QPOs. Their frequency increases similarly with \sa. In Fig.
\ref{bf_vs_hf}, we show the relation between LFQPO and upper HFQPO
frequencies when both QPOs are detected simultaneously and above
4~$\sigma$.  The correlation between the two frequencies is obvious.
It can be approximated with a power law of the form
$\nu_{LFQPO}=\alpha\nu_{HFQPO}^{\beta}$.  The first attempt of fitting
the data provided a high reduced \ch~because of the point representing
the July 15th observation lying off from the fit. We thus made the fit
taking out this point. We obtained $\alpha=(6\pm10)\times10^{-4}$ and
$\beta=1.6\pm0.3$ (reduced \ch~of 1.1) as parameters of the power law
drawn as the solid line on Fig.  \ref{bf_vs_hf}.

\subsection{Search for  oscillations in bursts}

For the bursts recorded in the 122~$\mu$s resolution mode, we have
computed thirty PDS of 1 second duration so that they covered the rising
and decaying parts of the bursts. Each was individually searched for
excess power.  No such excesses were found either in individual PDS
nor in their sum. We therefore conclude that no QPOs were found in
those bursts. We have set upper limits on the RMS, assuming a 3~$\sigma$ signal of FWHM 1~Hz in the PDS. We derive a 3~$\sigma$ upper
limit of 3.2~\% for the October 29th burst, and of 2.8~\% for the August
16th burst.

For the burst catcher mode, the situation is slightly different. The
data cover only the first 3.75 seconds of the bursts, and the Nyquist
frequency is  256~Hz which is below the typical frequency of a
burst oscillation \cite{vdk99}. Note however that coherent
oscillations were found at 7.6~Hz in a burst from Aql X-1
\cite{aql:schoelkopf91apj}.  For the burst that occured on May 5th, no
significant signals were found in PDS computed for different segments
durations (ranging from 1.25 to 3.75 seconds).  We have derived a 3~$\sigma$ upper limit of 2.8~\% (FWHM=1~Hz) for the 3.75 seconds burst
duration.

For the June 1st burst, we have also computed an FFT from the 3.75
seconds data set.  When the PDS so computed is rebinned by a factor of
4 ($\Delta \nu=1.1$~Hz), two signals are detected by our algorithm;
117.5~Hz (5.4~$\sigma$, RMS=$3.5\pm0.4$~\%), 225.7~Hz (5.0~$\sigma$,
RMS=$2.8\pm0.2$~\%). However, in the case of these PDS, the
corresponding significance detection level (taking into account the
number of trials; see section \ref{sec:search_technique}) is
only 85~\%, which is insufficient to claim a detection.

\section{Discussion}

\subsection{The nature of \xb}
The classification introduced by Hasinger and Van der Klis
\cite*{hasinger89aa} distinguishes two classes of LMXBs, the Z and the
atoll sources.  We list the main evidence suggesting that \xb~is an
atoll source.  First, \xb~has always been detected at X-ray
luminosities below $\sim1.5 \times 10^{37}$~\ergs, which is typical
for atoll sources.  The Z sources are roughly 10 times brighter.
Second, the orbital period of \xb~(50 min) is among the shortest
observed for LMXBs.  Now, it is observed that Z sources have longer
periods (more than 10 h) than atoll sources, consistent with the idea
that their companions are giants or subgiants \cite{vdk95xrb}.
Furthermore, almost all persistent dippers and bursters belong to the
atoll class (Cyg X-2 and GX 17+2 are the only Z sources to display
type I bursts \cite{kahn84apj,cygx2:smale98apjl,gx17:sztajno86mnras}).
Finally, no Z track in the color-color diagram has been observed
neither in the GINGA data \cite{yoshida92thesis}, nor in the RXTE
data. Hence, this favors the idea that \xb~belongs to the atoll class,
as suggested by Yoshida \cite*{yoshida92thesis}.

\subsection{The states of \xb}

Identifying the different atoll states in \xb~is not straightforward.
From the hardness-intensity diagrams, we can see that the emission, on
average, varies along the observations from a HS to a LS that could
correspond to banana and island states respectively.  However, we do
not observe clearly separated regions in the color-color diagram but
rather a continuous banana shape.  Furthermore, in its LS, \xb~does
not display HFN which is the typical aperiodic variability of atoll
sources in their island state; \xb~displays VLFN in its both states.
This behaviour was observed in the GINGA data as well. Unlike Yoshida
\cite*{yoshida92thesis} who associated the two regimes with island and
banana states, we believe that, during our observations, the source
was not in its island state; it probably remained on the banana branch
and hence never reached the state which may be at even lower
luminosity. Such a state may have been observed by BeppoSAX
\cite{1916:church98aa} since extrapolating the spectral parameters
reported by Church et al. \cite*{1916:church98aa} into the 2-50~keV
energy range gives a luminosity of $4.5\times10^{36}$~\ergs~while the
luminosity derived from the RXTE observations in the same energy band
ranges between $5\times10^{36}$ and
$1.4\times10^{37}$~\ergs~\cite{1916:bloser00apj}.

The properties and presence of HFQPOs strongly depend on source
states. For atoll sources, HFQPOs seem to occur at intermediate
inferred accretion rates: they are generally not observed in extreme
island or upper banana states corresponding respectively to the lowest
and highest accretion rates in a given source, as in e.g. 4U1608-52
\cite{1608:mendez99apjl}. Twin simultaneous HFQPOs have now been
detected in all atoll sources showing kHz variability but Aql X-1
\cite{vdk99}. They are always seen during banana states with the
exception of 4U1728-34
\cite{1728:ford98apjl,1728:stro96apjl,1728:mendez99apjl} and 4U1735-44
\cite{1735:ford98apjl}.  This gives further support to the idea that
\xb~was in a banana state during our observation. The highest
intensity state where no HFQPO is detected (February and March
observations) was therefore more likely the upper banana state.  The
lowest intensity state was likely the lower banana state, although no
HFQPO is detected during May 18th to 22nd observations where the count
rate is the lowest; maybe due to a lack of sensitivity (the RMS upper
limits corresponding to these observations and quoted in Table
\ref{tabqpo} are not really constraining).

\subsection{Tracking the accretion rate with \sa}

It has been shown that the relation between HFQPO frequency and
intensity could be much more complex than a roughly one to one
correlation as for example in 4U1820-30 \cite{1820:smale97apjl}.
Numerous branches are clearly visible in the HFQPO frequency versus
intensity diagram made with an extensive set of data from 4U1608-52
and 4U1728-34 \cite{1608:mendez99apjl,1728:mendez99apjl}.  Zhang et
al.  \cite*{aql:zhang98apjlb} also showed that Aql X-1, observed in
very distinct flux or count rate ranges, could display HFQPOs in the
same frequency range. The absence of a simple correlation had
previously been noticed in 4U1705-44 \cite{1705:ford98apjl}.  For
4U1608-52, the HFQPO frequency was reported to correlate well with the
count rate but only on timescales of hours \cite{1608:mendez99apjl}.
Thus, the relation between HFQPO frequency and intensity seems to
differ from source to source; sometimes being complex.  On the
contrary, the position on color-color diagram yields a more universal
and unique relation as in, e.g.  4U1608-52, 4U1728-34
\cite{1608:mendez99apjl,1728:mendez99apjl}. Our study of \xb~confirms
these results. The HFQPO frequency has a simpler relation with the
position \sa~on the color-color than with the count rate (Fig.
\ref{nu_hf_bf_vs_cts} and \ref{nu_hf_bf_vs_sa}). The frequency is well
correlated with \sa~within its full range.

Now, currently admitted models for HFQPOs, which are beat frequency
models (BFM), involve a bright spot at a specified radius of the
Keplerian disk, e.g. the sonic radius in the ``sonic-point model''
\cite{miller98apja} to produce the upper HFQPOs.  As the accretion
rate increases, the disk inner radius moves inwards until the
innermost stable orbit is reached.  Then, the HFQPO frequency is
supposed to correlate with the accretion rate up to a saturation.
Thus, the observed complex relations between HFQPO frequency and count
rate show that the count rate is not as good as \sa~to track the
accretion rate. This confirms the studies carried out with EXOSAT data
which revealed the position on color-color diagrams as a better
indicator of accretion rate than intensity
\cite{hasinger89aa,1636:vdk90apjl,cygx2:hasinger90aa}.

However, the position on the color-color diagram may not be the only
or the best indicator of the accretion rate or (and) of the timing
behaviour.  The properties of the HFQPO RMS amplitude found in
\xb~(Fig. \ref{nu_hf_bf_vs_cts} and \ref{nu_hf_bf_vs_sa}) may be new
evidence for this. Indeed, in the sonic-point model, the HFQPO
amplitude is expected to decrease as the accretion rate increases: the
optical depth and electron density increase near the neutron star, and
the oscillations generated there are more attenuated during their
propagation \cite{miller98apja}. Therefore, one would expect the HFQPO
RMS to be anticorrelated with the accretion rate. However, for \xb,
the HFQPOs RMS does not seem to be anticorrelated with \sa~but rather
may be anticorrelated with the count rate (Fig. \ref{nu_hf_bf_vs_sa}
and \ref{nu_hf_bf_vs_cts}).  Thus, the timing properties do not
probably depend only on the position \sa. Furthermore, it is not clear
why the HFQPOs RMS amplitude and frequency would be driven by
different parameters.

Moreover, we have shown that \xb~can display different timing
behaviours (presence or absence of HFQPOs) during observations
overlapping in the color-color diagram. The same situation occurs in
other sources, e.g. 4U1636-53 \cite{mendez98texas}.

One explanation may be that these timing behaviours are in fact the
same but seem different in appearance because of instrumental
limitations. For example, a QPO may be present in two observations
overlapping in the color-color diagram but not detected in one of them
because one of the detectors has been switched off leading to a loss
of sensitivity (see e.g. the RMS upper limits in Table \ref{tabqpo}).

A second explanation may be that within the timespan for which \sa~is
computed, the source might have moved in and out the \sa~span, washing
out temporal signals.


Another explanation is that the source has actually intrinsically
different timing behaviours. In this case, this is evidence that the
position in the color-color diagram is not the best indicator of the
timing behaviour of a source. Then two different conclusions may be
considered. First, assuming that the timing behaviour is governed by
the accretion rate, then \sa~is not a good indicator of the accretion
rate. This quantity may be better represented by a combination of
parameters or by other parameters than \sa. Second, assuming that
\sa~is a good indicator of the accretion rate, then the timing
behaviour is not governed only by the accretion rate.

\subsection{Low and high frequency quasi-periodic oscillations}
\label{sec:lfhf}

LFQPOs between a few and $\sim$~80~Hz have now been observed in
several atoll sources (see Table \ref{qporeview}). In 4U0614+09
\cite{ford97thesis}, 4U1608-52 \cite{1608:yu97apjl}, 4U1702-43
\cite{1702:markwardt99apjl}, 4U1728-34
\cite{1728:stro96apjl,1728:ford98apjl}, KS1731-260
\cite{1731:wijnands97apjl} and 4U1735-444 \cite{1735:wijnands98apjl},
such LFQPOs are detected simultaneously with HFQPOs. In the case of
4U0614+09, 4U1728-34 and 4U1702-43, the LFQPO frequency is observed to
increase with the HFQPO frequency and with the intensity of the
source. First, this indicates that the LFQPOs detected in atoll
sources are comparable to the horizontal branch QPOs (HBOs) detected
in Z sources, suggesting that similar physical processes are at work
in both kind of LMXBs.  Second, this indicates that LF and HFQPOs are
produced either by related mechanisms or by the same mechanism seen
under different aspects or occuring at different locations in the
disk.  Our study of \xb~confirms both these conclusions. For the first
time, we show a LFQPO in an atoll source varying in frequency and RMS
amplitude as a function of the position on the color-color diagram
(Fig.  \ref{nu_hf_bf_vs_sa}) the same way as HBOs in Z sources. We
note however that in \xb, the LFQPOs are not strictly correlated with
the source intensity, which is a characteristic of HBOs and also of
the atoll sources mentioned above showing LFQPOs and HFQPOs. \xb's
behaviour is thus different in this respect. This may be explained by
the fact that in \xb, the intensity is not correlated with the
position in the color-color diagram over the entire range where LFQPOs
are detected. On the contrary, the intensity is strictly correlated
with the position in the color-color diagram in horizontal branches
\cite{vdk95xrb} and probably in some portions of the track of atoll
sources.  In \xb, we clearly show (Fig.  \ref{nu_hf_bf_vs_sa} and
\ref{nu_hf_bf_vs_cts}) that \sa, and not the count rate, is the key
parameter of the quasi-periodic variability at low frequencies.  Thus,
the HF and LFQPOs frequencies depend the same way upon \sa.

Psaltis et al. \cite*{psaltis99apja} have shown that the frequency of
the QPOs detected between 0.1 and 100~Hz in numerous Z, atoll or black
hole binaries, simultaneously with HFQPOs followed one of a small
number of correlations with the lower HFQPO frequency. For \xb, we
could detect simultaneously the LFQPO and the lower HFQPO only during
three segments.  On the diagram derived by Psaltis et al.  \cite*[see
their figure 2]{psaltis99apja}, these points would fall in the region
where the two main correlations merge. Hence, our RXTE observation
does not help to conclude about the unified picture suggested for
QPOs.

The fact that HBOs and HFQPOs have now been observed simultaneously in
Z sources, as in Cyg X-2 \cite{cygx2:wijnands98apjl} calls into
question the models proposed for QPOs. Indeed, the magnetospheric BFM
was first proposed to explain the HBOs in Z sources
\cite{alpar85nature}.  Later, the same model was suggested to explain
the HFQPOs \cite{1728:stro96apjl} observed both in atoll and Z
sources. But simultaneous LFQPOs and HFQPOs are difficult to 
explain with the same model. This requires that the disk is still
present inside the radius where the magnetosphere couples the disk and
channels the matter \cite{miller98apja}. In this case, HBOs and HFQPOs
can be explained both with the magnetospheric BFM. Psaltis et al.
\cite*{psaltis99apjb} have shown that the correlations observed
between HBO and HFQPO frequencies in Z sources were indeed consistent
with this model.  However, this result has been shown so far for Z
sources only.  The striking similarities between the LFQPOs detected
in \xb~and HBOs suggest that both phenomena could be interpreted
within a same model.  But the magnetospheric BFM may not be easily
extrapolated to atoll sources because of their lower inferred magnetic
fields and accretion rates.

Another possible interpretation comes from Stella and Vietri
\cite*{stella98apjl} who proposed that the LFQPOs observed around
15-50~Hz in several atoll and Z (horizontal branch) sources could
result from the precession of the innermost disk regions. In the first
version of this model, the LFQPO frequency is the nodal precession
($\nu_{nod}$) of slightly titled orbits in these regions.  This
precession frequency (relativistic -Lense-Thirring- and classical
contributions) depends on the equation of state of the neutron star
but also on its spin frequency and on the Keplerian frequency of the
innermost accretion disk region. In the framework of BFMs, these
latter frequencies are inferred for LMXBs displaying twin HFQPOs, from
the frequency separation of the peaks and the frequency of the upper
peak respectively.  The precession frequency is predicted to vary
approximately as the square of the innermost Keplerian frequency.  For
\xb, the LFQPO frequency varies as $\nu_{K}^{1.6}$ where $\nu_K$ is
the inferred Keplerian frequency (see Fig. \ref{bf_vs_hf}). This is
not far from the expected quadratic dependence within the errors.
However, the observed LFQPO frequency is $\sim$~2 times greater than
the expected precession frequency drawn as a dashed line on Fig.
\ref{bf_vs_hf}.  We assumed a neutron star spin frequency of 348~Hz
which is the mean peak separation between the twin HFQPOs detected in
\xb.  Furthermore, we used the same parameters as Stella and Vietri
\cite*{stella98apjl} for the neutron star: mass $ M= 1.97 {\rm
  M}_{\odot}$, ratio $I_{45}/M = 1.98$ where $I_{45}$ is the moment of
inertia in units of 10$^{45}$ g cm$^2$.  With these values, Stella and
Vietri \cite*{stella98apjl} found that the precession frequency
matched the LFQPO frequency for three different atoll sources
(4U1728-34, 4U0614-09 and KS1731-260). Since the neutron star
parameters used yield precession frequencies close to the maximum
values that can be reached keeping the parameters in the ranges
allowed in classical neutron star models, it is unlikely that the
Lense-Thiring model can be pushed up to match the LFQPOs in \xb.  The
precession frequencies are actually lower than the observed LFQPOs
frequencies for any values of M and $I_{45}/M $ within their
reasonable respective ranges. This would also be the case if the spin
frequency were lower than the one inferred from the twin peaks
separation (348~Hz). On the opposite, the precession frequencies would
roughly match the LFQPOs if the spin frequency were twice that value
($\sim$~700~Hz). With a spin frequency of 348~Hz, matching the
observed LFQPO frequencies with precession frequencies would require a
ratio $I_{45}/M $ of $\sim$~3.5 which is too large. Indeed, this ratio
is inferred to be in the range $\sim$~0.5-2 for realistic rotating
neutron star models
\cite{stella98apjl,markovic98apj,kalogera99phys,miller98apjb}.

Similar conclusions have been reached for 4U1728-34 \cite[ from a
larger set of data than the one used by Stella and Vietri
\cite*{stella98apjl}]{1728:ford98apjl}, 4U1735-44
\cite{1735:wijnands98apjl}, 4U1702-42 \cite[in this case, the required
ratio $I_{45}/M $ is only 2.3 which is closer of the accepted range
than for other sources]{1702:markwardt99apjl} and for the Z sources GX
17+2, GX 5-1, GX 340+0, Cyg X-2 and Sco X-1
\cite{stella98apjl,gx340:jonker98apjl,psaltis99apjb,kalogera99phys}.
For the atoll source 4U1608-52, the precession frequency model does
not seem to apply either: the spin frequency predicted from the
Lense-Thirring model \cite{1608:yu97apjl} is inconsistent with the one
inferred from the frequency separation of the twin HFQPOs later
observed \cite{1608:mendez98apjla,1608:mendez98apjlb}.  In summary,
the LFQPOs roughly follow the dependence upon the Keplerian frequency
expected within the Lense-Thirring model but occur at a frequency
lower than the precession frequency for a growing group of binaries
that now includes \xb.  Thus, in order to explain the LFQPOs with the
Lense-Thirring effect, further hypothesis are required
\cite{markovic98apj,morsink99apj,armitage99apj,kalogera99phys,schaab99mnras}.
Another possibility is that the LFQPOs correspond to the second
harmonics of the precession frequency 2$\nu_{nod}$ instead of
$\nu_{nod}$ as proposed in a second version of the Lense-Thirring
model, maybe because of a modulation at twice the precession frequency
generated at the two points where the inclined orbit of the blobs
intersects the disk
\cite{vietri98apj,morsink99apj,stella98apjl,stella99apjl}.

However, Psaltis et al.  \cite*{psaltis99apjb} and Schaab and Weigel
\cite*{schaab99mnras} noticed another discrepancy with the model: the
LFQPO frequency does not follow the expected quadratic dependence on
the Keplerian frequency when the latter frequency is greater than 850
Hz in Z sources.  In addition, radiation forces in the disk could
significantly affect the precession frequencies and thus possibly
challenge the Lense-Thirring model as the explanation for the LFQPOs
\cite{miller99apj}.  Thus, the LFQPOs origins and relations with
HFQPOs remain unclear. 

Another model proposes to interpret both the LF and HF QPOs within the
same framework: the two-oscillator model
\cite{osherovich99apjl,tita99apjl,tita99apjlb}.  In this model, the
lower kHz QPO is the Keplerian frequency ($\nu_K$~$\sim$~700~Hz) at
the outer edge of the boundary layer between the Keplerian disk and
the neutron star. Blobs thrown out from this region into the
magetosphere oscillate in both radial and perpendicular modes to
produce an upper kHz QPO (at $\nu_h$~$\sim$~1000~Hz ) and a low
frequency QPO (at $\nu_L$~$\sim$~50~Hz) respectively.  Furthermore,
viscous and diffusive processes within the boundary layer produce
respectively a low frequency QPO (at $\nu_V$~$\sim$~30~Hz) and a break
(at $\nu_b$~$\sim$~8~Hz) in the PDS. In this model, the angle $\delta$
between the rotational frequency $\nu=\Omega/2\pi$ of the
magnetosphere and the normal to the neutron star disk is not zero.

The various observed QPOs of Sco X-1, 4U1728-34 and 4U1702-42 have
been succesfully identified in the framework of the two-oscillator
model \cite{tita99apjlb,1702:osherovich99apjl}. The angle $\delta$
derived is $8.3^\circ\pm1.0^\circ$, $5.5^\circ\pm0.5^\circ$ and
$3.9^\circ\pm0.2^\circ$ respectively for each source. Furthermore, the
predicted variation of $\nu_b$ as $\nu_V^{1.6}$ has been checked for
number of atoll and Z sources \cite{tita99apjlb}.

For \xb, the identification of the observed QPOs is not
straightforward. \xb~does not show any break in its PDS. Only one
LFQPO is detected in the frequency range where QPOs are predicted to
be present at $\nu_V$, $\nu_L$ and 2$\nu_L$.

We tried however to test the two-oscillator model.  $\Omega$, $\nu_K$
and $\nu_h$ follow the relation $\nu_h^2=\nu_K^2+(\Omega/\pi)^2$ (see
e.g., equation 2 in Osherovich and Titarchuk
\cite*{1702:osherovich99apjl}), so that the rotational frequency can
be derived when twin HFQPOs are detected simultaneously and
interpreted as $\nu_K$ and $\nu_h$.

\begin{figure}
\resizebox{\hsize}{!}{\includegraphics{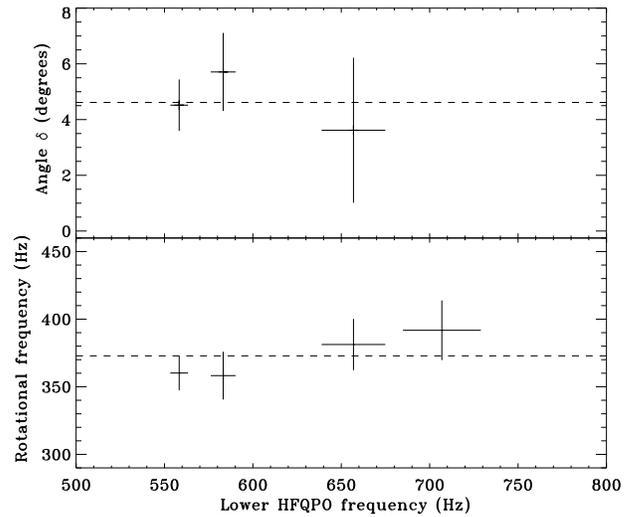}}
\caption{Inferred rotational frequency (bottom panel) and angle
  $\delta$ (top panel) as a function of the lower HFQPO frequency
  within the two-oscillator model. The dashed lines represent the mean
  values.  The four points in the bottom panel are inferred from the 4
  detections of twin HFQPOs above 550~Hz. The three points in the top
  panel correspond to the three cases where a LFQPO is simultaneously
  detected.}
\label{tita}
\end{figure}

The bottom panel of Fig. \ref{tita} shows the inferred rotational
frequency as a function of the lower HFQPO frequency, for the four
segments where twin HFQPOs are detected above 550~Hz (see Table
\ref{qporeview}).  Theoretically, $\Omega$ depends on the magnetic
structure of the neutron star's magnetosphere. In the case of \xb,
there are not enough simultaneous detections of $\nu_K$ and $\nu_h$ to
reconstruct the $\Omega$ profile. We thus use the approximation
$\Omega=\Omega_0=const$, as was done also for 4U1702-42
\cite{1702:osherovich99apjl}. As shown in Fig. \ref{tita} (bottom
panel), the rotational frequency is indeed consistent with being
constant with $\nu=\Omega_0/2\pi= 373\pm36$~Hz. We note that the pair
of QPOs at 224 and 514~Hz would correspond to a rotational frequency
of $231.2\pm1.3$~Hz inconsistent with the previous value and not
considered here.

Knowing $\Omega$, the angle $\delta$ can be derived when $\nu_K$ and
$\nu_L$ are measured simultaneously (see e.g. equation 5 in Osherovich
and Titarchuk \cite*{1702:osherovich99apjl}). The top panel of Fig.
\ref{tita} shows the inferred angle $\delta$ for \xb, assuming that
the LFQPO frequency is $\nu_L$. $\delta$ is consistent with being
constant with a mean value $\delta=4.6^\circ\pm3.1^\circ$. It is in
the range of values found for the other sources, but the error is
particularly large.

We note however that the relation between $\Omega$, $\nu_K$ and
$\nu_h$ written above requires $\nu_h>2\nu$, which becomes
$\nu_h>746$~Hz in the case of \xb. Now, according to Fig.
\ref{nu_hf_bf_vs_sa} (left) showing the HFQPO frequency as a function
of \sa, when only one HFQPO is detected, it seems to be the upper
HFQPO. If we assume that it is actually the case and that its
frequency is $\nu_h$, then the observed frequency range of $\nu_h$
would begin at $\sim$~500~Hz. This would be inconsistent with the
condition $\nu_h>746$~Hz derived according to the assumptions
considered above.

\subsection{Implications for the neutron star in \xb}

Observations of HFQPOs allow to derive constraints on the neutron star
present in the system. In the framework of beat frequency models, the
frequency difference between twin HFQPOs is interpreted as the spin
frequency of the neutron star. For \xb, the mean frequency separation
is 348~Hz for four segments of observations. This is consistent with
values reported in other LMXBs and would imply a neutron star rotating
with a 2.8 ms period. However, a frequency separation of 290~Hz is
detected for a pair of QPOs at 514 and 224~Hz. The spin frequency
detection at 348~Hz needs to be firmly confirmed.

\section{Conclusions}

We have analyzed the 1996 RXTE data of the X-ray burster and dipper
\xb. We confirm that it is an atoll source that was probably in its
lower and upper banana branch during the RXTE observations.

We report the discovery of both LFQPOs (5-80~Hz) and HFQPOs
(200-1300~Hz) and weak evidence for a 0.2~Hz QPO.  We show a
correlation between the LFQPOs and the HFQPOs frequencies suggesting
that both QPOs are produced either by related mechanisms or by the
same mechanism seen under different aspects or occuring at different
locations in the disk.  Furthermore, the LF and HFQPOs frequencies
both positively correlate with the position \sa~of the source in the
color-color diagram that is a key parameter of the timing behaviour
though probably not the only one.

Four twin HFQPOs are detected above 550~Hz with a frequency separation
consistent with being constant ($348\pm12$~Hz) suggesting a spin
frequency of 2.8 ms for the neutron star. A pair of QPOs below 550~Hz
is also detected with a frequency separation of 290~Hz.

Many more observations have been performed by RXTE.  They should help
to confirm the picture of the aperiodic variability discovered from
\xb. Especially, the 0.2~Hz QPO and signatures of the neutron star
rotation should be searched for in the persistent, dipping and
bursting emissions.

\begin{acknowledgements}
  This research has made use of data obtained through the High Energy
  Astrophysics Science Archive Research Center Online Service,
  provided by the NASA-Goddard Space Flight Center.  We thank E. Ford
  and M. Méndez for helpful comments. We are deeply endebted to an
  anonymous referee for its suitable and detailed comments that helped
  to improve this paper.
\end{acknowledgements}


\end{document}